\documentclass[aps,prd,twocolumn,superscriptaddress,nofootinbib,longbibliography,floatfix]{revtex4-2}
\usepackage{amsmath,amssymb,graphicx,bm,xcolor,textcomp}
\usepackage[colorlinks=true,linkcolor=blue,citecolor=blue,urlcolor=blue]{hyperref}

\newcommand{\bas}{\bar{\alpha}_S}
\def\beq{\begin{equation}}
\def\eeq{\end{equation}}
\def\bea{\begin{eqnarray}}
\def\eea{\end{eqnarray}}
\newcommand{\nn}{\nonumber}
\newcommand{\Lb}{\left(}
\newcommand{\Rb}{\right)}
\newcommand{\h}{\frac{1}{2}}
\def\pom{{I\!\!P}}
\newcommand{\eq}[1]{Eq.~(\ref{#1})}
\newcommand{\intl}{\int\limits}
\newcommand{\bg}{\bar{\gamma}}

\newcommand{\verificado}{}

\begin{document}

\title{KNO functions for dipole-dipole scattering from large Pomeron loops:\\ log-normal tails and a test in exactly solvable models}

\author{M. Roa}
\email{michael.roa@upla.cl}

\affiliation{Departamento de F\'isica, Universidad T\'ecnica Federico Santa Mar\'ia,
Casilla 110-V, Valpara\'iso, Chile}
\affiliation{Facultad de Ingenier\'ia, Laboratorio DataScience, Universidad de Playa
Ancha, ~~\\
 Leopoldo Carvallo 270, Valpara\'iso, Chile~~~~~~ ~~\\
}


\begin{abstract}
The sum of large Pomeron loops in QCD, based on the $t$-channel unitarity and on the solution of the Balitsky-Kovchegov (BK) equation in the saturation region, has been combined with the AGK (Abramovsky, Gribov and Kancheli) cutting rules to obtain multiplicity distributions with KNO (Koba, Nielsen and Olesen) scaling and an entropy equal to the logarithm of the gluon structure function. We reexamine these distributions for the BK amplitude and for dipole-dipole scattering. We show that the KNO functions that follow from the Gaussian representation of the asymptotic series have log-normal and not Gaussian tails, $\ln\Psi_{\rm BK}\simeq-\frac{1}{2}\ln^2\xi$ and $\ln\Psi_{\rm dd}\simeq-\frac{1}{4}\ln^2\xi$, and we give steepest-descent expressions for them in closed form in terms of the Lambert function. These functions are log-normal mixtures of the distributions of one and of two BFKL cascades; the summation is not unique, since the underlying moment problem is indeterminate. The KNO scaling requires that power-like factors in the coefficients of the series be discarded, and the alternative contour representation leads to functions that change sign. We then test the approach in two exactly solvable models in zero transverse dimensions. In the unitary toy model we confirm that the resummed series reproduces the exact $S$-matrix to within $2.5$--$10\%$, while the $S$-matrix itself changes by many orders of magnitude; this is the check on which the summation procedure rests. Nevertheless, the multiplicity distribution obtained from the same series with the AGK rules agrees with the exact distribution of the number of inelastic dipole-dipole interactions only as long as both cascades are dilute. At higher energies the exact mean grows as a power of rapidity, the KNO scaling is absent, and the entropy in the rest frame of the target grows as one half of the logarithm of the mean multiplicity. The amplitude is determined by rare configurations of the cascades with a small number of dipoles and the multiplicity by the typical ones, so that the agreement of the amplitudes does not validate the multiplicity distributions.
\end{abstract}

\maketitle

\section{Introduction}
\label{sec:intro}

The multiplicity distribution of gluons produced in high-energy collisions has become a central observable of small-$x$ QCD since it was proposed in Ref.~\cite{Kharzeev:2017qzs} that deep inelastic scattering (DIS) probes a maximally entangled state of the proton, whose entanglement entropy is fixed by the gluon structure function,
\beq \label{eq:KL}
S_E \,=\, \ln\!\big[ xG\Lb x,Q^2\Rb\big] .
\eeq
This relation has been confronted with the H1 data~\cite{Kharzeev:2021yyf,Hentschinski:2022rsa}, extended to hadronic collisions~\cite{Gotsman:2020bjc}, and studied in several models~\cite{Kutak:2023cwg,Braun:2026tga}. In the saturation region, where the Balitsky-Kovchegov (BK) equation~\cite{Balitsky:1995ub,Kovchegov:1999yj} replaces the linear BFKL evolution~\cite{Kuraev:1977fs,Balitsky:1978ic}, the multiplicity distribution of the produced gluons was found to be of the Koba-Nielsen-Olesen (KNO) type~\cite{Koba:1972ng}, with an entropy that depends on the approximations made~\cite{Levin:2023mwl,Contreras:2026gvm}.

A separate line of work has addressed the scattering of two dilute systems (dipole-dipole scattering), for which the BK equation is not sufficient and the Pomeron loops have to be summed. In a series of papers~\cite{Levin:2024eas,Levin:2024ire,Levin:2025oad,Levin:2025gbh,Levin:2025cgd,Levin:2026oaz} the sum of large Pomeron loops was performed using the $t$-channel unitarity in the form of Mueller, Patel, Salam and Iancu~\cite{Mueller:1996te,Iancu:2003zr}, with the dipole densities $\rho_n$ reconstructed from the solution to the BK equation deep in the saturation region~\cite{Levin:1999mw}. The scattering amplitude is then an asymptotic series of multi-Pomeron exchanges, whose coefficients are known only at large $n$ and grow as $\exp\Lb n^2/2\Rb$~\cite{Levin:2025cgd}; such a series cannot be summed by the Borel procedure. Applying the Abramovsky-Gribov-Kancheli (AGK) cutting rules~\cite{Abramovsky:1973fm} to each term of this series, Refs.~\cite{Levin:2024ire,Levin:2025cgd} obtained the multiplicity distributions of produced gluons both for the BK amplitude and for dipole-dipole scattering. In both cases the distribution shows KNO scaling,
\beq \label{eq:KNOdef}
\sigma_n \,=\, \frac{1}{N\Lb z\Rb}\,\Psi\Lb \xi\Rb, \qquad \xi \,=\,\frac{n}{N\Lb z\Rb},
\eeq
with a mean multiplicity proportional to $N\Lb z\Rb$, and hence to the Green's function of the BFKL Pomeron, and therefore with the entropy of \eq{eq:KL}.

Two features of these results motivate the present paper. First, the two available calculations use different procedures for summing the asymptotic series and lead to different KNO functions. Reference~\cite{Levin:2025cgd} finds $\Psi_{\rm dd}\Lb\xi\Rb = \sqrt{\Psi_{\rm BK}\Lb\xi\Rb}$ at large $\xi$, while Ref.~\cite{Levin:2024ire} and the one-dimensional models~\cite{Mueller:1996te,Levin:2024wtl} give a relation of the form $\Psi_{\rm dd}\Lb\xi\Rb = \Psi_{\rm BK}\Lb\sqrt{\xi}\Rb$. The origin of this difference was attributed in Ref.~\cite{Levin:2025cgd} to the summation procedure, but which of the two forms is correct has not been established. Second, the summation procedure of Ref.~\cite{Levin:2025cgd} is supported by the fact that it reproduces the exact scattering amplitude in the unitary toy model (UTM) of Ref.~\cite{Kovner:2024tin}, a zero-transverse-dimension model which satisfies both $t$- and $s$-channel unitarity. That check, however, concerns the amplitude. Whether the multiplicity distributions obtained from the same series through the AGK cutting rules are reproduced as well has not been tested.

In this paper we address both questions. Our results are the following.

(i) We reexamine the KNO functions of Ref.~\cite{Levin:2025cgd} and find that the integrals which define them do not have the Gaussian tails quoted there. A steepest-descent evaluation in terms of the Lambert $W$ function, confirmed by direct numerical integration, gives log-normal tails: $\ln\Psi_{\rm BK} \simeq -\h\ln^2\xi$ and $\ln\Psi_{\rm dd}\simeq -\tfrac{1}{4}\ln^2\xi$ at $\xi\gg1$. The relation $\Psi_{\rm dd}\simeq\sqrt{\Psi_{\rm BK}}$ holds only asymptotically and is not satisfied at $\xi\lesssim 10$. We also recalculate the constants that enter the entropy.

(ii) We show that the KNO scaling of Refs.~\cite{Levin:2024ire,Levin:2025cgd} relies on discarding the power-like factors of $1/n$ in the coefficients of the series, whose form at small $n$ is not known; with these factors kept, the distribution does not scale. We also show that the summation procedure of Ref.~\cite{Levin:2024ire} leads to a function $\Psi_{\rm BK}\Lb\xi\Rb$ which changes sign at $\xi\gtrsim1$ and therefore cannot be interpreted as a probability distribution.

(iii) We show that the KNO functions of Ref.~\cite{Levin:2025cgd} are log-normal mixtures of the multiplicity distributions of one and of two BFKL cascades, and we give all their moments. This representation also shows that the sum of the series is not unique even among positive functions, since the factor $\exp\Lb n^2/2\Rb$ defines an indeterminate moment problem.

(iv) We confront the approach with two exactly solvable models in zero transverse dimensions: the parton cascade with dipole merging of Ref.~\cite{Levin:2024mva} and the UTM~\cite{Kovner:2024tin,Levin:2024wtl}. In both models the distribution of the number of inelastic dipole-dipole interactions can be computed directly from the exact dipole probabilities, without any reference to an asymptotic series. In the UTM, where the resummed series reproduces the exact amplitude at all rapidities, the multiplicity distribution obtained from the AGK cutting rules agrees with the exact one only in a limited region: in the frame in which the rapidity is shared equally between the two dipoles, and as long as both cascades are dilute. Outside this region the exact mean grows as a power of rapidity instead of exponentially, the distribution does not show KNO scaling, its entropy in the target rest frame grows as $\h\ln\langle k\rangle$ instead of $\ln\langle k\rangle$, and the number of inelastic interactions depends on the frame.

(v) We trace this behavior to its origin. The AGK distribution derived from the resummed series is the distribution of the number of inelastic interactions between two dilute cascades. The amplitude at high energy is determined by rare configurations with far fewer dipoles than the typical ones, whose probabilities are little affected by saturation, while the multiplicity is determined by the typical configurations, which are saturated. In the language of the multi-Pomeron series, the exact eigenfunctions that describe the exchange of $k$ Pomerons vanish linearly, and not as the $k$-th power, when the scattering amplitude of a single dipole goes to zero, so that all multi-Pomeron exchanges contribute to the mean multiplicity.

The paper is organized as follows. In Sec.~\ref{sec:method} we recall the method of Refs.~\cite{Levin:2024ire,Levin:2025cgd}. Section~\ref{sec:qcd} contains the corrected asymptotics of the KNO functions in QCD, and Sec.~\ref{sec:assumptions} discusses the two assumptions behind the KNO scaling. Sections~\ref{sec:kappa} and~\ref{sec:utm} are devoted to the two exactly solvable models, and Sec.~\ref{sec:origin} to the origin of the discrepancy found in the second of them. The implications for QCD are discussed in Sec.~\ref{sec:discussion}, and Sec.~\ref{sec:conclusions} summarizes our conclusions. The appendices contain the details of the steepest-descent evaluation (Appendix~\ref{app:lambert}), of the contour representation (Appendix~\ref{app:contour}), the exact relations used in the two models (Appendix~\ref{app:exact}) and the numerical methods (Appendix~\ref{app:numerics}).

\section{Multiplicity distributions from the multi-Pomeron series: the method}
\label{sec:method}

In this section we recall the approach of Refs.~\cite{Levin:2024ire,Levin:2025cgd} and fix the notation. Nothing in this section is new.

\subsection{Scattering amplitude as a series of multi-Pomeron exchanges}

The starting point is the $t$-channel unitarity for the dipole-dipole scattering amplitude~\cite{Mueller:1996te,Iancu:2003zr,Levin:2005au},
\bea \label{eq:MPSI}
&& N\Lb Y, r, r' ; \vec{b}\Rb \,=\,\sum^\infty_{n=1}\Lb -1\Rb^{n+1} n!\intl \prod^n_{i=1} d^2 r_i\,d^2 r'_i\,d^2 b'_i\,d^2 \delta b_i \nn\\
&&\times\,\gamma^{BA}\Lb r_i,r'_i, \delta \vec{b}_i\Rb\,\rho_n\Lb Y - Y_0; \{ \vec{r}_i,\vec{b}_i\}\Rb\,\rho_n\Lb Y_0; \{ \vec{r}'_i,\vec{b}'_i\}\Rb ,\nn\\
\eea
where $\rho_n$ are the densities of $n$ dipoles in the wave functions of the two colliding dipoles of sizes $r$ and $r'$, evolved to rapidities $Y-Y_0$ and $Y_0$, and $\gamma^{BA}$ is the dipole-dipole amplitude in the Born approximation. The amplitude must not depend on $Y_0$.

Deep in the saturation region the amplitudes depend on a single variable~\cite{Levin:1999mw},
\beq \label{eq:z}
z\,=\,\bas\, \kappa \,Y \,+\,\xi_{r,r'}, \qquad \kappa \,\equiv\, \frac{\chi\Lb \gamma_{cr}\Rb}{1 - \gamma_{cr}}\,=\, - \frac{d \chi\Lb \gamma_{cr}\Rb}{d \gamma_{cr}},
\eeq
where $\bas\,\chi\Lb\gamma\Rb$ is the BFKL kernel in the anomalous-dimension representation and $\xi_{r,r'}$ is the logarithm of the ratio of the dipole sizes. In this region the Green's function of the BFKL Pomeron takes the form
\beq \label{eq:GP}
G_{\pom}\Lb z \Rb \,=\, N_0\, e^{\bg \,z}, \qquad \bg \,=\, 1 - \gamma_{cr},
\eeq
with a constant $N_0$. For the full leading-order kernel $\gamma_{cr}\approx 0.37$ and $\kappa\approx 4.88$. Reference~\cite{Levin:2025cgd} uses the leading-twist kernel, for which $\bg=\h$ and $\kappa=4$.

The BK equation sums the fan diagrams of the BFKL Pomeron calculus~\cite{Gribov:1983ivg,Braun:2000wr}. The amplitude of the scattering of a dipole with a dense target can therefore be written as
\beq \label{eq:fan}
N_{\rm BK}\Lb z\Rb \,=\, \sum_{n=1}^\infty C_n\, (-1)^{n-1}\, G^n_{\pom}\Lb z\Rb ,
\eeq
and the coefficients $C_n$, which determine the dipole densities $\rho_n = C_n \prod_i G_{\pom}\Lb z_i\Rb$, are fixed by the requirement that \eq{eq:fan} reproduces the solution of the BK equation deep in the saturation region~\cite{Levin:1999mw}. For the leading-twist kernel this gives, at large $n$~\cite{Levin:2025cgd},
\beq \label{eq:Cn}
C_n \,=\, \frac{1}{n}\,\exp\Lb \h\, n^2 \,-\, 2\,n\Rb .
\eeq
Substituting the densities into \eq{eq:MPSI}, with each Pomeron attached to $\gamma^{BA}$ normalized as $G_{\pom}/N_0$ as in Ref.~\cite{Levin:2025cgd}, and using the $t$-channel unitarity for the exchange of a single Pomeron, one obtains for the dipole-dipole $S$-matrix
\beq \label{eq:Sdd}
S_{\rm dd}\Lb z\Rb \,=\, 1 - N_{\rm dd}\Lb z\Rb \,=\, \sum^\infty_{n=0} \Lb - 1\Rb^n\, n! \,C^2_n\,\Lb \frac{G_{\pom}\Lb z\Rb}{N^2_0}\Rb^n .
\eeq
Both \eq{eq:fan} and \eq{eq:Sdd} are asymptotic series. Because of the factor $\exp\Lb n^2/2\Rb$ in \eq{eq:Cn} they are not Borel summable. The coefficients are known only at large $n$, and Ref.~\cite{Levin:2025cgd} emphasizes that different summation procedures can therefore lead to different results.

\subsection{AGK cutting rules and the multiplicity distribution}

The multiplicity distribution is obtained in two steps~\cite{Mueller:1996te,Levin:2024ire}. First, the $s$-channel unitarity for a single BFKL Pomeron, $2\,{\rm Im}\,G_{\pom}\Lb z\Rb = \sigma^{\rm BFKL}_{in}\Lb z\Rb$, identifies the cut Pomeron with the production of gluons which have a Poisson distribution with mean multiplicity $\bar n_{\pom} = \bg\, z$. Second, both amplitudes are alternating series of multi-Pomeron exchanges, $N\Lb z\Rb=\sum_{k=1}^{\infty}(-1)^{k+1}F_k\Lb z\Rb$, where $F_k>0$ is the absolute value of the contribution of the exchange of $k$ Pomerons. From \eq{eq:fan} and \eq{eq:Sdd},
\beq \label{eq:Fk}
F^{\rm BK}_k\Lb z\Rb \,=\, C_k\, G^k_{\pom}\Lb z\Rb , \qquad
F^{\rm dd}_k\Lb z\Rb \,=\, k!\,C^2_k\,\Lb \frac{G_{\pom}\Lb z\Rb}{N^2_0}\Rb^k .
\eeq
The AGK cutting rules~\cite{Abramovsky:1973fm} then give the cross section for cutting $n\geq1$ of them,

\beq \label{eq:AGK}
\sigma^k_n \,=\, (-1)^{k-n}\,\frac{k!}{(k - n)!\,n!}\,2^{k}\, F_k , \qquad
\sigma^{\rm AGK}_n \,=\,\sum_{k=n}^{\infty}\sigma^k_n .
\eeq
The cross section for the production of $m$ gluons is the convolution
\beq \label{eq:conv}
\sigma_m\Lb z \Rb \,=\,\sum_{n} \sigma_n^{\rm AGK}\Lb z \Rb\, \frac{\Lb n \, \bg \,z\Rb^m}{m!}\, e^{ -n\, \bg\,z} .
\eeq
At large $z$ the Poisson factor is a narrow function of $n$ around $n=m/(\bg z)$, while $\sigma^{\rm AGK}_n$ varies on the scale $n\sim G_{\pom}\Lb z\Rb/N_0$. The distribution of the produced gluons is then given by the distribution of the cut Pomerons with $n \to m/(\bg z)$, and in what follows we discuss $\sigma_n^{\rm AGK}$, omitting the superscript.

\subsection{Two procedures for summing the series}

\subsubsection{Contour representation}

In Ref.~\cite{Levin:2024ire} the BK $S$-matrix is represented as
\bea \label{eq:old}
S_{\rm BK}\Lb z\Rb &=& C\intl^{\epsilon + i \infty}_{\epsilon - i \infty} \frac{d \omega}{2\,\pi\,i}\, e^{ \frac{\bg^2\kappa}{2}\omega^2}\nn\\
&&\times \sum^\infty_{n = 0} \frac{(-1)^n}{n!}\, \frac{\Gamma\Lb \omega +n\Rb}{\Gamma\Lb \omega \Rb}\Lb \frac{G_{\pom}\Lb z\Rb}{N_0}\Rb^n ,
\eea
where the sum over $n$ is performed under the integral. Applying \eq{eq:AGK} to each term, summing over $k$ and taking $z\gg1$ with $n \sim G_{\pom}/N_0$, Ref.~\cite{Levin:2024ire} arrives at a KNO distribution in the variable $\xi = n N_0/\Lb 2 G_{\pom}\Lb z\Rb\Rb$,
\beq \label{eq:Psiold}
\Psi^{(\omega)}_{\rm BK}\Lb\xi\Rb \,=\, C\, e^{-\xi}\, F\Lb\xi\Rb ,\quad
F\Lb\xi\Rb \,=\intl^{\epsilon + i \infty}_{\epsilon - i \infty} \frac{d \omega}{2\,\pi\,i}\, \frac{e^{ \frac{\bg^2\kappa}{2}\omega^2}}{\Gamma\Lb \omega\Rb}\, \xi^{\,\omega} .
\eeq
The superscript $(\omega)$ labels this summation procedure.

\subsubsection{Gaussian representation}

In Ref.~\cite{Levin:2025cgd} the factor $\exp\Lb n^2/2\Rb$ of \eq{eq:Cn} is written as a Gaussian integral over an auxiliary variable $\lambda$, which makes the dependence of each term on $n$ exponential and allows the sum over $n$ to be performed under the integral. The factor $2^k$ of \eq{eq:AGK} and the factor $e^{-2k}$ of $C_k$ (squared in the dipole-dipole case) combine with $G_{\pom}$ into the scale variables
\beq \label{eq:N}
N_{\rm BK}\Lb z\Rb \,=\, 2\,N_0\, e^{\h\Lb z - 4\Rb} , \qquad
N_{\rm dd}\Lb z\Rb \,=\, \frac{2}{N_0}\, e^{\h\Lb z - 8\Rb} ,
\eeq
which follow from \eq{eq:Fk} with $\bg=\h$. In what follows $N\Lb z\Rb$ stands for $N_{\rm BK}\Lb z\Rb$ in the BK case and for $N_{\rm dd}\Lb z\Rb$ in the dipole-dipole case. The constant $N_0$ only fixes the scale of $N\Lb z\Rb$ and does not enter the KNO functions, which depend on $n$ through $\xi = n/N\Lb z\Rb$. Reference~\cite{Levin:2025cgd} uses $N_0\, e^{\h\Lb z-4\Rb}$ in both cases, which differs from \eq{eq:N} by constant factors.

With this notation the AGK rules give for the BK amplitude, with the factor $1/n$ of \eq{eq:Cn} omitted,
\beq \label{eq:sBK}
\sigma_n \,=\, \frac{1}{\sqrt{2\pi}} \intl^{\infty}_{-\infty} d \lambda \, e^{ - \h \lambda^2}\,\frac{ \Lb N\Lb z\Rb e^{-\lambda}\Rb^n}{\Lb 1+ N\Lb z\Rb e^{ - \lambda }\Rb^{n+1}} ,
\eeq
and for dipole-dipole scattering, with the factor $1/n^2$ omitted,
\bea
\sigma_n &=& \frac{1}{2\sqrt{\pi}}\intl^{\infty}_{-\infty} d \lambda \,e^{ - \frac{\lambda^2}{4}} \nn\\
&&\times\intl^{\infty}_0 dt \,e^{-t}\,\frac{\Lb t\,e^{\lambda} N\Lb z\Rb\Rb^n}{\Lb 1+ t\,e^{\lambda} N\Lb z\Rb\Rb^{n+1}} , \label{eq:sdd}
\eea
where the integral over $t$ represents the factor $n!$ of \eq{eq:Sdd}. Reference~\cite{Levin:2025cgd} argues that these factors are not valid at small $n$ and discards them. We return to this point in Sec.~\ref{sec:assumptions}.

If the factors are kept, the sum over $k$ in \eq{eq:AGK} can still be performed exactly. With $y$ the variable raised to the power $k$ and $V \equiv y/\Lb 1+y\Rb$, one has
\bea
\sum_{k=n}^{\infty}\Lb -1\Rb^{k-n}\binom{k}{n}\frac{y^k}{k} &=& \frac{V^n}{n} , \nn\\
\sum_{k=n}^{\infty}\Lb -1\Rb^{k-n}\binom{k}{n}\frac{y^k}{k^2} &=& \frac{1}{n}\sum_{m=n}^{\infty}\frac{V^m}{m} . \label{eq:sums}
\eea
For the BK amplitude, with $y = N\Lb z\Rb e^{-\lambda}$, this gives
\beq \label{eq:sBK1n}
\sigma_n \,=\, \frac{1}{\sqrt{2\pi}}\,\frac{1}{n} \intl^{\infty}_{-\infty} d \lambda \, e^{ - \h \lambda^2}\,\Lb\frac{ N\Lb z\Rb e^{-\lambda}}{1+ N\Lb z\Rb e^{ - \lambda }}\Rb^{n} ,
\eeq
and for dipole-dipole scattering, with $y = t\,e^{\lambda} N\Lb z\Rb$,
\bea
\sigma_n &=& \frac{1}{2\sqrt{\pi}}\,\frac{1}{n}\intl^{\infty}_{-\infty} d \lambda \,e^{ - \frac{\lambda^2}{4}} \nn\\
&&\times\intl^{\infty}_0 dt \,e^{-t}\sum_{m=n}^{\infty}\frac{V^m}{m} . \label{eq:sdd1n}
\eea
Reference~\cite{Levin:2025cgd} gives for the BK case $1/n$ times the integrand of \eq{eq:sBK}. This differs from \eq{eq:sBK1n} by the factor $1+N\Lb z\Rb e^{-\lambda}$ under the integral: it follows from \eq{eq:sBK} multiplied by $1/n$, but not from the AGK rules applied to the series with the coefficients of \eq{eq:Cn}.

In the limit $N\Lb z\Rb\to\infty$ at fixed $\xi = n/N\Lb z\Rb$, Eqs.~(\ref{eq:sBK}) and (\ref{eq:sdd}) take the KNO form of \eq{eq:KNOdef} with
\bea
\Psi_{\rm BK}\Lb\xi\Rb &\propto& \intl^{\infty}_{-\infty} d \lambda \; \exp\Lb - \h \lambda^2 +\lambda -\xi\, e^{\lambda}\Rb , \label{eq:PsiBK}\\
\Psi_{\rm dd}\Lb\xi\Rb &\propto& \intl^{\infty}_{-\infty} d \lambda \; e^{- \frac{\lambda^2}{4} - \lambda}\, K_0\Lb 2 \sqrt{ \xi\, e^{-\lambda}}\Rb , \label{eq:Psidd}
\eea
where $K_0$ is the modified Bessel function. Equations~(\ref{eq:Psiold}), (\ref{eq:PsiBK}) and (\ref{eq:Psidd}) are the objects studied in the next two sections.

\section{KNO functions in QCD: corrected asymptotics}
\label{sec:qcd}

In this section we evaluate the KNO functions of \eq{eq:PsiBK} and \eq{eq:Psidd}. We normalize them as probability densities in $\xi$, $\int_0^\infty d\xi\,\Psi\Lb\xi\Rb = 1$. The integrals over $\xi$ can be taken under the integral over $\lambda$, which gives\verificado
\bea
\Psi_{\rm BK}\Lb\xi\Rb &=& \frac{1}{\sqrt{2\pi}}\intl^{\infty}_{-\infty} d \lambda \; \exp\Lb - \h \lambda^2 +\lambda -\xi\, e^{\lambda}\Rb , \label{eq:PsiBKn}\\
\Psi_{\rm dd}\Lb\xi\Rb &=& \frac{1}{\sqrt{\pi}}\intl^{\infty}_{-\infty} d \lambda \; e^{- \frac{\lambda^2}{4} - \lambda}\, K_0\Lb 2 \sqrt{ \xi\, e^{-\lambda}}\Rb , \label{eq:Psiddn}
\eea
with the first moments\verificado
\beq \label{eq:means}
\langle \xi\rangle_{\rm BK} \,=\, e^{1/2}, \qquad \langle \xi\rangle_{\rm dd} \,=\, e .
\eeq
The mean multiplicities are therefore $\bar n_{\rm BK} = e^{1/2} N_{\rm BK}\Lb z\Rb$ and $\bar n_{\rm dd} = e\, N_{\rm dd}\Lb z\Rb$: both are proportional to $G_{\pom}\Lb z\Rb$ and have the same dependence on $z$, while their ratio is a constant that depends on the normalization $N_0$.

\subsection{Dipole-target scattering}

\begin{figure*}[tb]
\includegraphics[width=0.95\textwidth]{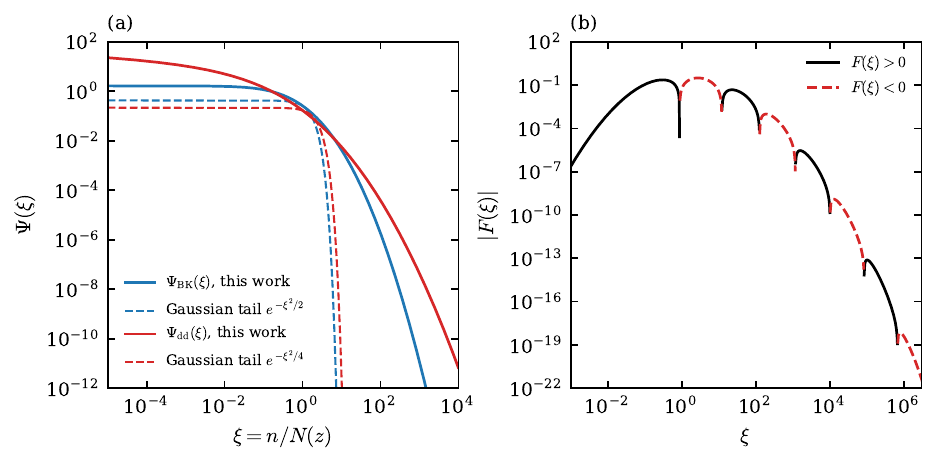}
\caption{\label{fig:qcd} (a) Normalized KNO functions of Eqs.~(\ref{eq:PsiBKn}) and (\ref{eq:Psiddn}) (solid lines), obtained with the Gaussian representation of Ref.~\cite{Levin:2025cgd}. The dashed lines show the Gaussian tails $\exp\Lb-\xi^2/2\Rb$ and $\exp\Lb-\xi^2/4\Rb$, normalized to the solid lines at $\xi=1$. (b) Absolute value of the function $F\Lb\xi\Rb$ of \eq{eq:Psiold}, obtained with the contour representation of Ref.~\cite{Levin:2024ire}, for $\bg^2\kappa/2 = 0.96$. Solid (dashed) segments correspond to $F>0$ ($F<0$); the dips are the zeros of $F$, at which it changes sign.}
\end{figure*}

The exponent of \eq{eq:PsiBKn}, $f\Lb\lambda\Rb = -\h\lambda^2+\lambda-\xi e^{\lambda}$, has a single maximum at
\beq \label{eq:spBK}
1-\lambda_{SP} \,=\, \xi\, e^{\lambda_{SP}} \,=\, W\Lb e\,\xi\Rb \,\equiv\, W ,
\eeq
where $W$ is the Lambert function. This is the saddle-point equation of Ref.~\cite{Levin:2025cgd}. At the saddle point $f = \h - W - \h W^2$ and $f'' = -\Lb 1+W\Rb$, so that\verificado
\beq \label{eq:PsiBKsd}
\Psi_{\rm BK}\Lb\xi\Rb \,\simeq\, \frac{1}{\sqrt{1+W}}\,\exp\Lb \h - W - \h W^2\Rb ,
\eeq
with $W = W\Lb e\,\xi\Rb$. Details are given in Appendix~\ref{app:lambert}. At $\xi\to0$ \eq{eq:PsiBKsd} reproduces the exact value $\Psi_{\rm BK}\Lb 0\Rb=e^{1/2}$. At $\xi \gg 1$, $W\Lb e\xi\Rb = \ln \xi - \ln\ln \xi + \dots$, and
\beq \label{eq:tailBK}
\ln \Psi_{\rm BK}\Lb\xi\Rb \,=\, -\h \ln^2 \xi \,+\, \ln\xi\,\ln\ln\xi \,+\, \mathcal{O}\Lb \ln \xi\Rb .
\eeq
The tail of the distribution is of the log-normal type. Table~\ref{tab:kno} and Fig.~\ref{fig:qcd}(a) show that \eq{eq:PsiBKsd} agrees with the numerical evaluation of \eq{eq:PsiBKn} within $0.01$ in $\ln\Psi$ for $1\leq\xi\leq10^6$.

Reference~\cite{Levin:2025cgd} states that $\lambda_{SP} = -\xi$ at large $\xi$ and concludes that $\Psi_{\rm BK}\propto \exp\Lb-\xi^2/2\Rb$. We find that this does not follow from \eq{eq:spBK}, which gives $\lambda_{SP} = 1 - W \simeq -\ln\xi$. The difference is not a matter of preasymptotic corrections: at $\xi=10^4$, for example, $\ln\Psi_{\rm BK} = -41.7$, while a Gaussian tail would give a value of the order of $-5\times10^{7}$.

\begin{table}[tb]
\caption{\label{tab:kno} Logarithms of the normalized KNO functions: numerical integration of Eqs.~(\ref{eq:PsiBKn}) and (\ref{eq:Psiddn}) compared with the steepest-descent results of Eqs.~(\ref{eq:PsiBKsd}) and (\ref{eq:Psiddsd}). The last column is the ratio $\ln\Psi_{\rm dd}/\ln\Psi_{\rm BK}$ of the numerical values, which equals $\h$ if $\Psi_{\rm dd}=\sqrt{\Psi_{\rm BK}}$.}
\begin{ruledtabular}
\begin{tabular}{rrrrrc}
$\xi$ & \multicolumn{2}{c}{$\ln\Psi_{\rm BK}$} & \multicolumn{2}{c}{$\ln\Psi_{\rm dd}$} & ratio\\
 & numerical & Eq.~(\ref{eq:PsiBKsd}) & numerical & Eq.~(\ref{eq:Psiddsd}) & \\
\hline
$1$ & $-1.35$ & $-1.35$ & $-1.78$ & $-1.71$ & 1.32\\
$10$ & $-5.46$ & $-5.46$ & $-5.18$ & $-5.14$ & 0.95\\
$10^2$ & $-13.21$ & $-13.22$ & $-10.16$ & $-10.13$ & 0.77\\
$10^3$ & $-25.18$ & $-25.18$ & $-16.95$ & $-16.93$ & 0.67\\
$10^4$ & $-41.65$ & $-41.66$ & $-25.73$ & $-25.72$ & 0.62\\
$10^6$ & $-88.81$ & $-88.81$ & $-49.70$ & $-49.69$ & 0.56\\
\end{tabular}
\end{ruledtabular}
\end{table}

\subsection{Dipole-dipole scattering}

For \eq{eq:Psiddn} we follow Ref.~\cite{Levin:2025cgd} and use the asymptotic form $K_0\Lb x\Rb \simeq \sqrt{\pi/(2x)}\,e^{-x}$. The exponent is then $g\Lb\lambda\Rb = -\frac{1}{4}\lambda^2-\frac{3}{4}\lambda-2\sqrt{\xi}\,e^{-\lambda/2}$, and the saddle point is determined by\verificado
\beq \label{eq:spdd}
\zeta\, e^{\zeta} \,=\, e^{3/4}\sqrt{\xi}, \qquad \lambda_{SP} \,=\, 2\,\zeta - \frac{3}{2} ,
\eeq
so that $\zeta = W\Lb e^{3/4}\sqrt{\xi}\Rb$. With $g\Lb\lambda_{SP}\Rb = -\zeta^2-2\zeta+\frac{9}{16}$ and $g''\Lb\lambda_{SP}\Rb = -\h\Lb 1+\zeta\Rb$ we obtain\verificado
\beq \label{eq:Psiddsd}
\Psi_{\rm dd}\Lb\xi\Rb \,\simeq\, \sqrt{\frac{\pi}{1+\zeta}}\;\xi^{-1/4}\exp\Lb -\zeta^2 - 2\zeta + \frac{9}{16}\Rb .
\eeq
At the saddle point the argument of $K_0$ equals $2\zeta$, so that \eq{eq:Psiddsd} requires $\zeta\gtrsim1$, i.e., $\xi\gtrsim 1$. At $\xi \gg1$, $\zeta \simeq \h\ln\xi$ and
\beq \label{eq:taildd}
\ln \Psi_{\rm dd}\Lb\xi\Rb \,=\, -\frac{1}{4} \ln^2 \xi \,\Lb 1 + \mathcal{O}\Lb \frac{\ln\ln\xi}{\ln \xi}\Rb\Rb .
\eeq
The comparison with the numerical integration, in which the exact function $K_0$ is used, is given in Table~\ref{tab:kno}.

Equations~(\ref{eq:spdd}) differ from the corresponding equations of Ref.~\cite{Levin:2025cgd}, which read $\zeta e^{\zeta} = \h e^{3/4}\sqrt{\xi}$ and $\lambda_{SP} = \frac{3}{2}+2\zeta$; the latter value of $\lambda_{SP}$ is not a stationary point of $g\Lb\lambda\Rb$. Evaluating the integrand at that point changes $\ln\Psi_{\rm dd}$ by $2$ to $4$ units in the range of Table~\ref{tab:kno}. As in the dipole-target case, the tail $\Psi_{\rm dd}\propto\exp\Lb-\xi^2/4\Rb$ quoted in Ref.~\cite{Levin:2025cgd} does not follow from \eq{eq:Psiddn}.

\subsection{The relation between $\Psi_{\rm dd}$ and $\Psi_{\rm BK}$}

Equations~(\ref{eq:tailBK}) and (\ref{eq:taildd}) show that $\ln\Psi_{\rm dd}/\ln\Psi_{\rm BK}\to\h$ at $\xi\to\infty$. In this sense the relation $\Psi_{\rm dd}\Lb\xi\Rb = \sqrt{\Psi_{\rm BK}\Lb\xi\Rb}$ of Ref.~\cite{Levin:2025cgd} holds, although for log-normal and not for Gaussian tails. However, the corrections to the leading terms are only logarithmically suppressed, and the limit is approached very slowly: the ratio is $0.95$ at $\xi=10$ and still $0.56$ at $\xi = 10^6$ (see Table~\ref{tab:kno}). We note that $\xi$ is defined with different scales in the two cases, $N_{\rm BK}$ and $N_{\rm dd}$ of \eq{eq:N}, whose ratio $N_{\rm BK}/N_{\rm dd} = N_0^2\, e^{2}$ depends on the normalization $N_0$. Comparing the two functions at the same multiplicity $n$ replaces $\xi$ by $N_0^2 e^{2}\,\xi$ in $\Psi_{\rm dd}$. This changes $\ln\Psi_{\rm dd}$ only at order $\ln\xi$ and leaves the limit $\ln\Psi_{\rm dd}/\ln\Psi_{\rm BK}\to\h$ unchanged, but the values of the ratio at finite $\xi$ in Table~\ref{tab:kno} depend on this choice.   At $\xi\lesssim 10$, where most of the probability is concentrated, the two functions are not related in this simple way. At small $\xi$ they are qualitatively different: $\Psi_{\rm BK}$ tends to a constant, while $\Psi_{\rm dd}$ has an integrable logarithmic singularity, $\Psi_{\rm dd}\propto\ln\Lb1/\xi\Rb$.

\subsection{Log-normal mixtures and moments}
\label{sec:mixture}

Equations~(\ref{eq:PsiBKn}) and (\ref{eq:Psiddn}) have a simple probabilistic structure, which explains the form of the tails and will be used in Secs.~\ref{sec:origin} and \ref{sec:discussion}. In \eq{eq:PsiBKn} the function $e^{\lambda}\exp\Lb-\xi e^{\lambda}\Rb$ is an exponential distribution of $\xi$ with the mean value $e^{-\lambda}$. The exponential distribution is the KNO function of the BFKL cascade, i.e., the large-multiplicity limit of the geometric distribution $N^n/\Lb1+N\Rb^{n+1}$~\cite{Mueller:1993rr,Mueller:1996te,Kharzeev:2017qzs}. Therefore
\beq \label{eq:mixBK}
\Psi_{\rm BK}\Lb\xi\Rb \,=\,\intl^{\infty}_{-\infty} \frac{d\lambda}{\sqrt{2\pi}}\; e^{-\h\lambda^2}\,\Big[\, e^{\lambda}\, e^{-\xi\, e^{\lambda}}\Big]
\eeq
is a mixture of the KNO functions of a single cascade, in which the logarithm of the mean multiplicity has a Gaussian distribution with unit variance. In the same way,
\beq \label{eq:mixdd}
\Psi_{\rm dd}\Lb\xi\Rb =\intl^{\infty}_{-\infty} \frac{d\lambda}{2\sqrt{\pi}}\; e^{-\frac{1}{4}\lambda^2}\,\Big[\, 2\,e^{-\lambda} K_0\Lb 2\sqrt{\xi\, e^{-\lambda}}\Rb\Big] ,
\eeq
where $2K_0\Lb2\sqrt{\xi}\Rb = \int dx\,dy\,e^{-x-y}\,\delta\Lb\xi-xy\Rb$ is the distribution of the product of two independent variables with exponential distributions, i.e., the KNO function for the number of pairs of dipoles in two independent cascades; the logarithm of its mean has a Gaussian distribution with variance $2$. The Gaussian weights are the images of the factors $\exp\Lb n^2/2\Rb$ and $\exp\Lb n^2\Rb$ of the coefficients $C_n$ and $C_n^2$ in Eqs.~(\ref{eq:fan}) and (\ref{eq:Sdd}).

Two consequences follow. First, all moments are known in closed form:\verificado
\beq \label{eq:moments}
\langle \xi^q\rangle_{\rm BK} = \Gamma\Lb q+1\Rb\, e^{q^2/2}, \qquad \langle \xi^q\rangle_{\rm dd} = \Gamma^2\Lb q+1\Rb\, e^{q^2} ,
\eeq
which contains \eq{eq:means}. The normalized moments $C_q=\langle\xi^q\rangle/\langle\xi\rangle^q$ are
\beq \label{eq:Cq}
C^{\rm BK}_q = q!\; e^{q(q-1)/2}, \qquad C^{\rm dd}_q = \Lb q!\Rb^2 e^{q(q-1)} ,
\eeq
so that $C_2^{\rm BK}=2e\approx5.4$ and $C_2^{\rm dd}=4e^2\approx29.6$, to be compared with $C_2=2$ and $C_2=4$ for one and two cascades without the Gaussian weight. The distributions are much broader than those of the BFKL cascade. Second, the tails of the mixtures are determined by the Gaussian weights and not by the exponential tails of the functions in the square brackets: a large value of $\xi$ is reached more easily by a fluctuation of the mean than by a fluctuation at fixed mean. This is the origin of the log-normal behavior of Eqs.~(\ref{eq:tailBK}) and (\ref{eq:taildd}), with the coefficients $1/(2\sigma^2)$ of $\ln^2\xi$ for the variances $\sigma^2=1$ and $2$.

\subsection{Entropy}

For a distribution with KNO scaling, $\mathcal{P}_n = \Psi\Lb n/N\Rb/N$, the entropy $S_E = -\sum_n \mathcal{P}_n\ln\mathcal{P}_n$ at large $N$ is
\beq \label{eq:entropy}
S_E \,=\, \ln \bar n \,+\, c, \qquad c \,=\, -\intl_0^\infty d\xi\,\Psi\Lb\xi\Rb\ln\Psi\Lb\xi\Rb \,-\,\ln\langle\xi\rangle .
\eeq
The leading term reproduces \eq{eq:KL} if $\bar n$ is identified with the gluon structure function, as in Refs.~\cite{Levin:2024ire,Levin:2025cgd}. For the constants we obtain numerically\verificado
\beq \label{eq:consts}
c_{\rm BK} \,=\, 0.81, \qquad c_{\rm dd} \,=\, 0.09 ,
\eeq
to be compared with $-1.706$ and $-0.217$ quoted in Ref.~\cite{Levin:2025cgd}. We have checked the value of $c_{\rm BK}$ by computing the entropy directly from \eq{eq:sBK}, without the factor $1/n$, at finite $N$: $S_E - \ln N = 1.360$, $1.323$ and $1.312$ for $N=30$, $100$ and $300$, which converges to $c_{\rm BK} + \h = 1.306$.

The results of this section correct the form of the KNO functions but do not change the leading behavior of the entropy obtained in Ref.~\cite{Levin:2025cgd}. Whether the KNO scaling itself is a reliable consequence of the approach is the subject of the following sections.

\section{Two assumptions behind the KNO scaling}
\label{sec:assumptions}

\subsection{The power-like factors in the coefficients $C_n$}

The KNO functions of Sec.~\ref{sec:qcd} were obtained from Eqs.~(\ref{eq:sBK}) and (\ref{eq:sdd}), in which the factor $1/n$ of \eq{eq:Cn}, and the factor $1/n^2$ in the dipole-dipole case, were discarded. We now show that this step is not a matter of normalization: it decides whether the distribution has KNO scaling at all.

The first moment can be computed exactly in all four cases, Eqs.~(\ref{eq:sBK}), (\ref{eq:sdd}), (\ref{eq:sBK1n}) and (\ref{eq:sdd1n}). Since $\sum_{n\geq1} n\, y^n/\Lb 1+y\Rb^{n+1} = \sum_{n\geq1} V^n = y$, and the same sum $y$ is obtained from the right-hand sides of \eq{eq:sums} multiplied by $n$, one has in all cases $\sum_n n\,\sigma_n = \langle\xi\rangle N\Lb z\Rb$, with $\langle\xi\rangle$ of \eq{eq:means}. The mean multiplicity is therefore
\beq \label{eq:nbarsin}
\bar n \,=\, \frac{\langle\xi\rangle\, N\Lb z\Rb}{\sigma_{in}}, \qquad \sigma_{in} \,=\, \sum_{n\geq1}\sigma_n ,
\eeq
and the four cases differ only in $\sigma_{in}$. Without the factors, $\sigma_{in}$ is the average of $V<1$ and tends to $1$ at large $N$. With them, the sum over $n$ gives $\sum_n V^n/n = \ln\Lb 1+y\Rb$ and $\sum_n n^{-1}\sum_{m\geq n}V^m/m = {\rm Li}_2\Lb V\Rb + \h\ln^2\Lb 1+y\Rb$, so that
\bea
\sigma^{\rm BK}_{in} &=& \frac{1}{\sqrt{2\pi}} \intl^{\infty}_{-\infty} d \lambda \, e^{ - \h \lambda^2}\,\ln\Lb 1+N\Lb z\Rb e^{-\lambda}\Rb \,\simeq\, \ln N , \nn\\
\sigma^{\rm dd}_{in} &=& \frac{1}{2\sqrt{\pi}}\intl^{\infty}_{-\infty} d \lambda \,e^{ - \frac{\lambda^2}{4}} \intl^{\infty}_0 dt \,e^{-t} \nn\\
&&\times\left[{\rm Li}_2\Lb V\Rb + \h\ln^2\Lb 1+y\Rb\right] \,\simeq\, \h\ln^2 N , \label{eq:sin1n}
\eea
with $y$ and $V$ as in \eq{eq:sdd1n}. In the BK case \eq{eq:sBK1n} gives at large $N$ the probabilities $\mathcal{P}_n \simeq \widetilde\Psi\Lb n/N\Rb/\Lb n\ln N\Rb$, with $\widetilde\Psi\Lb\xi\Rb = \int_\xi^\infty d\xi'\,\Psi_{\rm BK}\Lb\xi'\Rb$ and $\widetilde\Psi\Lb 0\Rb = 1$. Therefore
\beq \label{eq:with1n}
\mathcal{P}_n \,\simeq\, \frac{1}{n\,\ln N}\quad \Lb n \ll N\Rb, \qquad \frac{\bar n}{N}\,\simeq\,\frac{e^{1/2}}{\ln N}\,\to\,0 ,
\eeq
and the entropy grows as $S_E \simeq \h\ln N + \ln\ln N$ instead of $\ln N$. The distribution is concentrated at $n\ll \bar n$, the ratio $\bar n/N$ vanishes at large $z$, and $\mathcal{P}_n$ is not a function of $n/\bar n$ only. For dipole-dipole scattering with the factor $1/n^2$ the effect is stronger: $\bar n/N \simeq 2e/\ln^2 N$.

Table~\ref{tab:factors} illustrates these statements with a direct evaluation of Eqs.~(\ref{eq:sBK}) and (\ref{eq:sBK1n}) at finite $N$, in which the sums over $n$ are performed numerically. Without the factor $1/n$, $\bar n/N$ and $S_E-\ln N$ approach the constants $e^{1/2}$ and $c_{\rm BK}+\h$ of Sec.~\ref{sec:qcd}. With it, $\bar n/N$ decreases as $e^{1/2}/\sigma^{\rm BK}_{in}$ and $S_E - \ln N$ decreases without bound. For dipole-dipole scattering with the factor $1/n^2$, Eqs.~(\ref{eq:nbarsin}) and (\ref{eq:sin1n}) give $\bar n/N = 0.379$, $0.237$ and $0.164$ for $N=30$, $100$ and $300$, while without it $\bar n/N > e$.

\begin{table}[tb]
\caption{\label{tab:factors} Mean multiplicity and entropy of the BK distribution at finite $N$, without [\eq{eq:sBK}] and with [\eq{eq:sBK1n}] the factor $1/n$.}
\begin{ruledtabular}
\begin{tabular}{rcccc}
 & \multicolumn{2}{c}{without $1/n$} & \multicolumn{2}{c}{with $1/n$}\\
$N$ & $\bar n/N$ & $S_E - \ln N$ & $\bar n/N$ & $S_E - \ln N$ \\
\hline
30 & 1.733 & 1.360 & 0.478 & $-0.26$\\
100 & 1.675 & 1.323 & 0.357 & $-0.75$\\
300 & 1.658 & 1.312 & 0.289 & $-1.20$\\
$\infty$ & 1.649 & 1.306 & 0 & $-\infty$\\
\end{tabular}
\end{ruledtabular}
\end{table}

The factors $1/n$ and $1/n^2$ change the tails of the KNO functions only by factors that vary as powers of $\xi$, up to logarithms, which are subleading with respect to the log-normal behavior of Eqs.~(\ref{eq:tailBK}) and (\ref{eq:taildd}). Their effect is at small $n$, where the coefficients $C_n$ are not known~\cite{Levin:2025cgd}. The KNO scaling, the mean multiplicity $\bar n\propto N\Lb z\Rb$ and the entropy $S_E = \ln N\Lb z\Rb$ are thus not consequences of the known large-$n$ behavior of $C_n$ alone: they hold under the additional assumption that the true coefficients have no power-like suppression at $n\lesssim N\Lb z\Rb$. We note that the same step is present in the contour representation: in passing to \eq{eq:Psiold} the ratio $\Gamma\Lb n+\omega\Rb/n!\simeq n^{\omega-1}$ is replaced by $n^{\omega}$.

\subsection{Positivity in the contour representation}

We now turn to the function $F\Lb\xi\Rb$ of \eq{eq:Psiold}. The integrand is an entire function of $\omega$, and the contour can be moved to ${\rm Re}\,\omega = c\equiv -\ln\xi/(2a)$, where $a = \bg^2\kappa/2$. This gives the exact representation\verificado
\bea \label{eq:Fstable}
F\Lb\xi\Rb &=& \exp\Lb -\frac{\ln^2\xi}{4\,a}\Rb R\Lb\xi\Rb ,\nn\\
R\Lb\xi\Rb &=& \intl^{\infty}_{-\infty}\frac{dt}{2\pi}\; e^{-a\,t^2}\;{\rm Re}\,\frac{1}{\Gamma\Lb c + i\,t\Rb} ,
\eea
which is free of numerical cancellations. At small $\xi$ one has $c>0$ and $F>0$. For $\xi>1$, however, $c<0$, and $1/\Gamma\Lb c+it\Rb$ oscillates as a function of $c$ with unit period and with an amplitude that grows as $\Gamma\Lb 1-c\Rb$. As a result $F\Lb\xi\Rb$ changes sign. For the leading-order values $\gamma_{cr}=0.3725$, $\kappa=4.883$ ($a=0.96$) the zeros are at
\beq \label{eq:zeros}
\xi \,=\, 0.874,\;\; 12.1,\;\; 126,\;\; 1.17\times10^{3},\;\; 1.0\times10^{4},\;\dots
\eeq
[see Fig.~\ref{fig:qcd}(b)], and the ratio of consecutive zeros decreases slowly toward $e^{2a}\approx 6.8$, which corresponds to unit spacing in $c$. In particular, $F<0$ for $0.874<\xi<12.1$. We have checked that the sign change is not produced by the limit $n\to\infty$: the expression of Ref.~\cite{Levin:2024ire} with $\Gamma\Lb n+\omega\Rb$ kept, evaluated at $2G_{\pom}/N_0 = 200$, is negative at $n=400$, $1000$ and $2000$.

The steepest-descent estimate of Ref.~\cite{Levin:2024ire} for the same integral contains, in the present notation, the factor $1/\Gamma\Lb c\Rb$, whose argument is negative for $\xi>1$; it changes sign as well, although this is not discussed there. We also note that \eq{eq:Fstable} gives the exponent $-\ln^2\xi/(2\bg^2\kappa)$, instead of $-\ln^2\xi/(2\kappa)$ quoted in Ref.~\cite{Levin:2024ire}.

For dipole-dipole scattering Ref.~\cite{Levin:2024ire} obtains $\sigma_n$ as a double contour integral over $\omega_1$ and $\omega_2$ and an integral over an auxiliary variable $t$, which originates from the factor $n!$ in \eq{eq:Sdd}. The two contour integrals factorize exactly at any $n$:\verificado
\bea \label{eq:sddold}
\sigma_n &=& C^2\intl_0^\infty \frac{dt}{t}\; e^{-t}\,F\Lb t\Rb\,H_n\Lb t\,X\Rb ,\nn\\
H_n\Lb y\Rb &=& \intl^{\epsilon + i \infty}_{\epsilon - i \infty} \frac{d \omega}{2\,\pi\,i}\; e^{a\,\omega^2}\,\frac{\Gamma\Lb n+\omega\Rb}{\Gamma\Lb\omega\Rb\,n!}\,\frac{y^{\,n}}{\Lb 1+y\Rb^{n+\omega}} ,
\eea
where $X \equiv 2\,G_{\pom}\Lb z\Rb/N_0^2$ and $F$ is the function of \eq{eq:Psiold}. In the KNO limit, $n\to\infty$ and $X\to\infty$ at fixed $\xi = n/X$, one has $\Gamma\Lb n+\omega\Rb/n!\to n^{\omega-1}$ and $y^n\Lb 1+y\Rb^{-n-\omega}\to y^{-\omega}e^{-n/y}$, so that $H_n\Lb tX\Rb \to n^{-1}e^{-\xi/t}F\Lb\xi/t\Rb$ and\verificado
\bea \label{eq:Psiddold}
n\,\sigma_n \,\to\,\Psi^{(\omega)}_{\rm dd}\Lb\xi\Rb &\propto&\intl_0^\infty \frac{dt}{t}\; e^{-t-\xi/t}\,F\Lb t\Rb\,F\Lb \xi/t\Rb \nn\\
&=& \intl_{-\infty}^{\infty} du\; e^{-e^{u}-\xi e^{-u}}\,F\Lb e^{u}\Rb\,F\Lb \xi\, e^{-u}\Rb .\nn\\
\eea
The measure $dt/t$ does not lead to a singularity at $t=0$: the factor $e^{-\xi/t}$ vanishes there faster than any power of $t$, while $F$ is a bounded function which vanishes both at small and at large values of its argument [see \eq{eq:Fstable} and Fig.~\ref{fig:qcd}(b)]. The second form of \eq{eq:Psiddold}, in the variable $u=\ln t$, makes this explicit, and also shows the symmetry of the integrand under $t\to\xi/t$. The integrand is concentrated in the region $|u|\lesssim 3$; for $\xi=1$ the integral changes by less than one part in $10^{8}$ when the limits are extended from $|u|\leq4$ to $|u|\leq16$.

Evaluating \eq{eq:Psiddold} numerically with \eq{eq:Fstable} we find that $\Psi^{(\omega)}_{\rm dd}$ is negative for $0.22<\xi<1.9$. To verify that this is not an artifact of the KNO limit, we have also computed $n\,\sigma_n$ directly from \eq{eq:sddold} at finite $n$ and $X$. The results, shown in Table~\ref{tab:eq32}, converge to \eq{eq:Psiddold} as $X$ increases and are negative at $\xi=0.5$ and $\xi=1$ for all values of $X$ considered.

\begin{table}[tb]
\caption{\label{tab:eq32} The quantity $n\,\sigma_n/C^2$ computed from \eq{eq:sddold} at $n=\xi X$, compared with its KNO limit, \eq{eq:Psiddold}, for $a=0.9616$. All entries are multiplied by $10^{3}$.}
\begin{ruledtabular}
\begin{tabular}{rrrrrr}
$\xi$ & $X=50$ & $X=200$ & $X=800$ & $X=3200$ & Eq.~(\ref{eq:Psiddold})\\
\hline
0.1 & 26.73 & 27.09 & 27.08 & 27.07 & 27.07\\
0.5 & $-12.38$ & $-12.95$ & $-13.08$ & $-13.12$ & $-13.13$\\
1.0 & $-6.84$ & $-6.74$ & $-6.72$ & $-6.71$ & $-6.71$\\
3.0 & 1.29 & 1.31 & 1.31 & 1.32 & 1.32\\
\end{tabular}
\end{ruledtabular}
\end{table}

A function which is negative in the region where most of the events are expected cannot be interpreted as a multiplicity distribution. We conclude that, of the two summation procedures, only the Gaussian representation of Ref.~\cite{Levin:2025cgd} leads to positive KNO functions. The relation $\Psi_{\rm dd}\Lb\xi\Rb = \Psi_{\rm BK}\Lb\sqrt{\xi}\Rb$, which follows from the large-$\xi$ estimates in the contour representation, is therefore not supported.

\section{Exact test I: parton cascade with dipole merging}
\label{sec:kappa}

The Gaussian representation of Sec.~\ref{sec:method} was applied in Ref.~\cite{Levin:2024mva} to a zero-transverse-dimension model in which the multiplicity distribution can be computed exactly. We use this model as a first test.

\subsection{The model and its exact solution}

The model is a parton cascade with two processes, the decay of one dipole into two and the merging of two dipoles into one. The probability $P_n\Lb Y\Rb$ to find $n$ dipoles at rapidity $Y$ obeys~\cite{Levin:2024mva}
\bea \label{eq:master}
\frac{d P_n}{d Y} &=& -n\,P_n + (n-1)\,P_{n-1} \nn\\
&+& \kappa\Lb - \frac{n (n -1)}{2}\,P_n + \frac{n (n + 1)}{2}\,P_{n+1}\Rb ,
\eea
where $Y$ is measured in units of the inverse Pomeron intercept, and $\kappa\ll1$ is the relative strength of the merging (we keep the notation of Ref.~\cite{Levin:2024mva}; this parameter is not related to $\kappa$ of \eq{eq:z}). The generating function $Z\Lb Y,u\Rb=\sum_n P_n\Lb Y\Rb u^n$ satisfies
\beq \label{eq:Zkappa}
\frac{ \partial Z}{\partial Y}\,=\,- u\Lb 1-u\Rb \frac{ \partial Z}{\partial u}\,+\,\h\,\kappa\, u \Lb 1 - u\Rb \frac{ \partial^2 Z}{\partial u^2} ,
\eeq
with $Z\Lb 0,u\Rb=u$ and $Z\Lb Y,1\Rb=1$, and the $S$-matrix for the scattering with a single dipole is $S\Lb Y\Rb = Z\Lb Y, e^{-\kappa}\Rb = \sum_n e^{-\kappa n}P_n\Lb Y\Rb$.

Equation~(\ref{eq:master}) is a finite-range linear system which we solve numerically with $n\leq1500$; the total probability is conserved to better than $10^{-5}$. At large $Y$ the splitting and the merging balance each other, and the distribution approaches a stationary one. The stationary solution of \eq{eq:Zkappa} with $Z\Lb 0\Rb=0$ and $Z\Lb 1\Rb=1$ is\verificado
\beq \label{eq:Zinf}
Z_{\infty}\Lb u \Rb \,=\,\frac{e^{2u/\kappa}-1}{e^{2/\kappa}-1} ,
\eeq
i.e., a Poisson distribution with mean $2/\kappa$, restricted to $n\geq1$. The expression given in Ref.~\cite{Levin:2024mva}, $\Lb 1-e^{-2u/\kappa}\Rb/\Lb 1-e^{-2/\kappa}\Rb$, differs from \eq{eq:Zinf} by the sign of the exponents and does not satisfy \eq{eq:Zkappa}. From \eq{eq:Zinf},
\beq \label{eq:Sinf}
S\Lb Y\to\infty\Rb \,=\, Z_\infty\Lb e^{-\kappa}\Rb \,\simeq\, e^{-2}\Lb 1+\kappa\Rb ,
\eeq
which is $0.139$ for $\kappa=0.03$. In this model the number of dipoles saturates at $n\simeq 2/\kappa$, and the $S$-matrix does not vanish at high energy.

\subsection{The resummed series versus the exact amplitude}

In Ref.~\cite{Levin:2024mva} the $S$-matrix is written as a series of multi-Pomeron exchanges with the Green's functions $G_{k\pom}=\exp\Lb\Lb k+\kappa k^2\Rb Y\Rb$,
\beq \label{eq:Skappa}
S\Lb Y\Rb =\sum^\infty_{k=0} \Lb - \kappa\Rb^k \Lb 1- \h \kappa \Lb k+2\Rb k \Lb k-1\Rb\Rb e^{ \Lb k + \kappa k^2\Rb Y} ,
\eeq
and the factor $\exp\Lb\kappa k^2 Y\Rb$ is represented as a Gaussian integral. We find that the identity needed for this purpose is\verificado
\bea \label{eq:gauss}
e^{ \Lb k + \kappa k^2\Rb Y} &=& e^{ \Lb 1 - \kappa\Rb k Y}\intl^\infty_{-\infty}
\frac{d \lambda}{ 2 \sqrt{\pi \kappa Y} }\nn\\
&&\times \exp\Lb - \frac{ \Lb \kappa Y + \lambda\Rb^2}{ 4 \kappa Y} - \lambda\,k\Rb ,
\eea
while Ref.~\cite{Levin:2024mva} has $\Lb\kappa Y-\lambda\Rb^2$ in the exponent, which corresponds to $\exp\Lb\Lb k+\kappa k^2-2\kappa k\Rb Y\Rb$. After the sum over $k$ is performed under the integral, the series is replaced by a function which vanishes at large $Y$, as $\exp\Lb-\Lb 1-3\kappa\Rb Y\Rb$ according to Ref.~\cite{Levin:2024mva}.

Table~\ref{tab:Skappa} compares the result with the exact $S$-matrix. The two agree at $Y\lesssim2$, where a few terms of \eq{eq:Skappa} are sufficient. At larger $Y$ the resummed series decreases to zero, while the exact $S$-matrix approaches the constant of \eq{eq:Sinf}. The correction of the Gaussian center in \eq{eq:gauss} does not change this conclusion.

\begin{table}[tb]
\caption{\label{tab:Skappa} $S$-matrix of the model of \eq{eq:master} for $\kappa=0.03$: exact solution compared with the resummed series of Ref.~\cite{Levin:2024mva}, as published and with the Gaussian center of \eq{eq:gauss}.}
\begin{ruledtabular}
\begin{tabular}{rccc}
$Y$ & exact & as published & with \eq{eq:gauss} \\
\hline
1 & 0.925 & 0.926 & 0.922\\
3 & 0.685 & 0.653 & 0.612\\
5 & 0.369 & 0.239 & 0.191\\
8 & 0.170 & 0.022 & 0.014\\
20 & 0.139 & 0.000 & 0.000\\
\end{tabular}
\end{ruledtabular}
\end{table}

We attribute the disagreement to the coefficients of the series rather than to the summation. The intercepts $k+\kappa k^2$ and the prefactors in \eq{eq:Skappa} are the first terms of an expansion in $\kappa$ and can be trusted only for $k\ll1/\kappa$, whereas the resummation uses them at all $k$. At large $Y$ the result is determined by the terms with $k\gtrsim1/\kappa$.

\subsection{Distribution of the cut dipoles}

In a zero-dimensional cascade the multiplicity distribution can be obtained without the multi-Pomeron series. In the rest frame of the target each of the $n$ dipoles of the projectile scatters independently, with the $S$-matrix $e^{-\kappa}$; the probability of an inelastic interaction of a given dipole is $p = 1-e^{-2\kappa}$. The cross section for exactly $k$ inelastic interactions is therefore
\beq \label{eq:cuts}
\sigma_k\Lb Y\Rb \,=\,\sum_{n\geq k} P_n\Lb Y\Rb \binom{n}{k}\,p^k\Lb 1-p\Rb^{n-k} ,
\eeq
which is the parton-model form of the AGK cutting rules~\cite{Kovchegov:1999ji,Levin:2024wtl}: $\sum_{k\geq1}\sigma_k = 1 - Z\Lb Y,e^{-2\kappa}\Rb$ is the inelastic cross section. We have checked \eq{eq:cuts} in the limit $\kappa\to0$ of \eq{eq:master} at fixed $p$, where $P_n$ is a geometric distribution with mean $N=e^Y$ and \eq{eq:cuts} gives $\sigma_k = \Lb1-1/N\Rb^{k-1}\Lb pN\Rb^k/\Lb pN+1-p\Rb^{k+1}$; at $N\gg1$ and $p\ll1$ this is the distribution of the BFKL cascade, $\Lb pN\Rb^k/\Lb1+pN\Rb^{k+1}$, which follows from the AGK rules for the eikonal series.

\begin{figure*}[tb]
\includegraphics[width=0.95\textwidth]{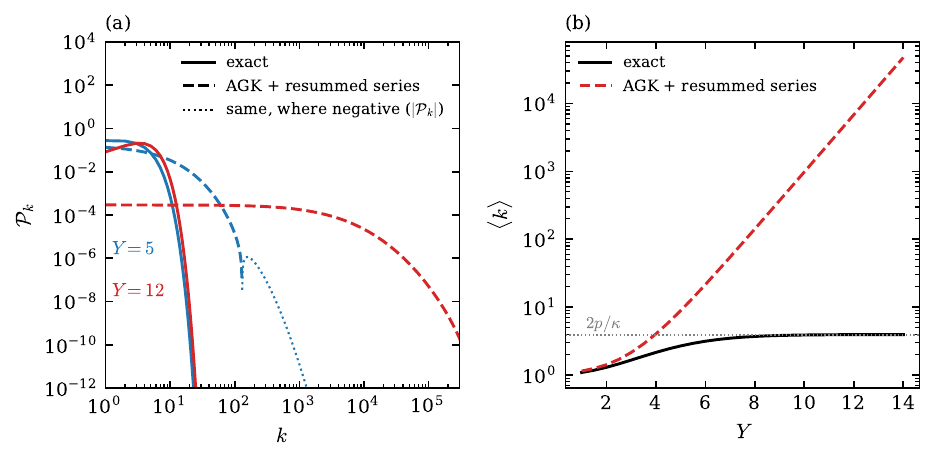}
\caption{\label{fig:kappa} Model of \eq{eq:master} with $\kappa=0.03$. (a) Distribution of the number $k$ of cut dipoles at $Y=5$ and $Y=12$. Solid lines: exact result, \eq{eq:cuts}. Dashed lines: AGK cutting rules applied to the resummed series, Ref.~\cite{Levin:2024mva}, normalized to the sum over all $k\geq1$. Dotted line: absolute value of the same expression in the region where it is negative. (b) Mean number of cut dipoles as a function of $Y$; the dotted horizontal line is the asymptotic value $2p/\kappa$ of the exact solution.}
\end{figure*}

\begin{table}[tb]
\caption{\label{tab:cuts} Mean number of cut dipoles for $\kappa=0.03$: exact result of \eq{eq:cuts}, with the entropy of the exact distribution, and the AGK result of Ref.~\cite{Levin:2024mva}.}
\begin{ruledtabular}
\begin{tabular}{rccc}
 & \multicolumn{2}{c}{exact} & AGK, resummed series\\
$Y$ & $\langle k\rangle$ & $S_E$ & $\langle k\rangle$ \\
\hline
2 & 1.30 & 0.71 & 1.43 \\
3 & 1.67 & 1.12 & 2.16 \\
5 & 2.70 & 1.72 & 8.91 \\
8 & 3.72 & 1.98 & 142 \\
12 & 3.95 & 2.01 & $6.8\times10^{3}$ \\
\end{tabular}
\end{ruledtabular}
\end{table}

The exact distribution is compared in Fig.~\ref{fig:kappa} and Table~\ref{tab:cuts} with the result of applying \eq{eq:AGK} to the terms of the resummed series~\cite{Levin:2024mva}\footnote{The second term of the corresponding expression of Ref.~\cite{Levin:2024mva} contains a variable $z$ which is not defined there; we read it as $\zeta = \exp\Lb\Lb1-\kappa\Rb Y-\lambda\Rb$.}. The exact distribution saturates: at $Y\gtrsim 8$ it no longer depends on $Y$, with $\langle k\rangle\simeq 2p/\kappa\simeq4$ and $S_E\simeq 2.0$. The AGK distribution keeps broadening with $Y$, and its mean grows as $2\kappa\exp\Lb\Lb1-\kappa\Rb Y\Rb$: it exceeds the exact mean by a factor of $3$ at $Y=5$ and by more than three orders of magnitude at $Y=12$ [Fig.~\ref{fig:kappa}(b)]. In addition, the AGK expression is negative at large $k$ for all values of $Y$ considered (at $k\geq130$ for $Y=5$), so that it does not define an entropy. The result $S_E = \Lb 1-\kappa\Rb Y$ of Ref.~\cite{Levin:2024mva} is not a property of the model, in which the entropy of the distribution of the cut dipoles is bounded.

This test shows that the procedure can fail for both the amplitude and the multiplicity distribution when the coefficients of the series are extrapolated beyond their range of validity. A more stringent test is provided by a model in which the resummed series does reproduce the exact amplitude.

\section{Exact test II: the unitary toy model}
\label{sec:utm}

\subsection{The model and the resummed amplitude}

The UTM~\cite{Mueller:1996te,Kovner:2024tin} is defined by the evolution equation for the probabilities to find $n$ dipoles,
\beq \label{eq:utm}
\frac{d P_n}{ d \tilde Y}=- \Lb 1-e^{- \gamma n}\Rb P_n +\Lb 1-e^{- \gamma (n - 1) }\Rb P_{n-1} ,
\eeq
with $P_n\Lb 0\Rb=\delta_{n1}$, where $\gamma\ll1$ is the amplitude for the scattering of two dipoles (not to be confused with the anomalous dimension of Sec.~\ref{sec:method}) and $\tilde Y$ is a rescaled rapidity, in which the exchange of a single Pomeron grows as $\exp\Lb\Lb e^{\gamma}-1\Rb\tilde Y\Rb$. The rate of emission of a new dipole is proportional to $n$ at $n\ll1/\gamma$, as in the BFKL cascade, and saturates at $n\gg1/\gamma$. The $S$-matrix,
\beq \label{eq:Sframe}
S\Lb\tilde Y\Rb =\sum_{n,m} e^{ - \gamma\, n\,m} \,P_n\Lb \tilde Y_0\Rb\,P_m\Lb \tilde Y - \tilde Y_0\Rb ,
\eeq
does not depend on $\tilde Y_0$: the model satisfies the $t$-channel unitarity exactly, and it satisfies the $s$-channel unitarity as well~\cite{Kovner:2024tin}. In the rest frame of the target, $\tilde Y_0=\tilde Y$, \eq{eq:Sframe} reduces to $S = Z\Lb e^{-\gamma},\tilde Y\Rb$ with $Z\Lb u,\tilde Y\Rb = \sum_n u^n P_n\Lb\tilde Y\Rb$.

The generating function has the expansion~\cite{Kovner:2024tin}
\bea
Z\Lb u, \tilde Y \Rb &=&\sum^\infty_{k=0} C_k\,\Phi_k\Lb u\Rb\, e^{ \Delta_k \tilde Y}, \nn\\
\Phi_k\Lb u\Rb &=&\prod^{k-1}_{l=0} \Lb e^{ - \gamma l} -\frac{1}{u}\Rb , \nn\\
C_k &=& e^{\gamma k (k+1)/2} , \qquad \Delta_k \,=\, e^{\gamma k}-1 . \label{eq:Zutm}
\eea
which is the series of multi-Pomeron exchanges of this model. The coefficients $C_k$ grow as a Gaussian in $k$, as in \eq{eq:Cn}, and the intercepts $\Delta_k$ grow exponentially. For small $\gamma$, $C_k\Phi_k\Lb e^{-\gamma}\Rb\simeq\Lb-\gamma\Rb^k k!$, and the series is summed in Refs.~\cite{Kovner:2024tin,Levin:2024wtl} by expanding $\exp\Lb\Delta_k\tilde Y\Rb = e^{-\tilde Y}\sum_j e^{\gamma k j}\,\tilde Y^j/j!$, which makes each term exponential in $k$, in the same way as the Gaussian representation of Sec.~\ref{sec:method}. The result is
\beq \label{eq:Sutm}
S\Lb \tilde Y\Rb = e^{- \tilde Y}\sum^\infty_{j=0} \frac{\tilde Y^j}{j!}\intl^\infty_0 d\tau\, \frac{e^{- \tau}}{ 1+\tau\,\gamma\,N_j} , \qquad N_j = e^{\gamma j} .
\eeq

We solve \eq{eq:utm} numerically, with the total probability conserved to better than $10^{-5}$. Table~\ref{tab:Sutm} confirms the statement of Ref.~\cite{Kovner:2024tin}: \eq{eq:Sutm} reproduces the exact $S$-matrix within $2.5\%$ for $\gamma=0.025$ and within about $10\%$ for $\gamma=0.1$, while $S$ changes by many orders of magnitude. This is the check of the summation procedure which Ref.~\cite{Levin:2025cgd} refers to.

\begin{table}[tb]
\caption{\label{tab:Sutm} $S$-matrix of the UTM: exact solution of \eq{eq:utm} and the resummed series, \eq{eq:Sutm}.}
\begin{ruledtabular}
\begin{tabular}{rllrll}
\multicolumn{3}{c}{$\gamma=0.025$} & \multicolumn{3}{c}{$\gamma=0.1$}\\
$\tilde Y$ & exact & resummed & $\tilde Y$ & exact & resummed\\
\hline
20 & 0.960 & 0.962 & 20 & 0.627 & 0.650\\
100 & 0.792 & 0.797 & 50 & 0.158 & 0.170\\
200 & 0.346 & 0.351 & 100 & $4.17\,10^{-3}$ & $4.55\,10^{-3}$\\
400 & $1.107\,10^{-2}$ & $1.131\,10^{-2}$ & 200 & $7.50\,10^{-7}$ & $8.25\,10^{-7}$\\
800 & $1.572\,10^{-6}$ & $1.610\,10^{-6}$ & 300 & $8.80\,10^{-11}$ & $9.69\,10^{-11}$\\
\end{tabular}
\end{ruledtabular}
\end{table}

\subsection{Multiplicity distribution}

Applying the AGK rules of \eq{eq:AGK} to the terms of the same series, Ref.~\cite{Levin:2024wtl} obtains
\beq \label{eq:sAGKutm}
\sigma^{\rm AGK}_n = e^{- \tilde Y}\sum^\infty_{j=0} \frac{\tilde Y^j}{j!}\intl^{\infty}_0 \!d \tau\, \frac{e^{-\tau}\Lb2\tau\gamma N_j\Rb^n}{\Lb1+2\tau\gamma N_j\Rb^{n+1}} ,
\eeq
which has the same structure as \eq{eq:sdd}. Its mean is\verificado
\beq \label{eq:meanAGK}
\langle k\rangle_{\rm AGK} = \frac{2\gamma}{\sigma_{in}^{\rm AGK}}\,\exp\Lb\Lb e^{\gamma}-1\Rb\tilde Y\Rb ,
\eeq
where $\sigma_{in}^{\rm AGK}=\sum_{n\geq1}\sigma_n^{\rm AGK}\leq 1$; i.e., it grows as the single-Pomeron exchange, as it must for any distribution obtained from the AGK rules.

As in Sec.~\ref{sec:kappa}, the number of inelastic dipole-dipole interactions can be computed from the exact probabilities. In a frame in which the projectile is evolved to $\tilde Y_0$ and the target to $\tilde Y-\tilde Y_0$, each of the $n\,m$ pairs of dipoles interacts inelastically with the probability $p=1-e^{-2\gamma}$, and
\beq \label{eq:cutsutm}
\sigma_k =\sum_{n,m} P_n\Lb \tilde Y_0\Rb P_m\Lb \tilde Y - \tilde Y_0\Rb \binom{n m}{k}\,p^k\Lb 1-p\Rb^{nm-k} .
\eeq
The sum over $k\geq1$ gives $1-S\big|_{\gamma\to2\gamma}$, the inelastic cross section. In the BFKL cascade this quantity reproduces the distribution of the cut Pomerons (Sec.~\ref{sec:kappa}). We consider the rest frame of the target, $\tilde Y_0=\tilde Y$, and the symmetric frame, $\tilde Y_0=\tilde Y/2$.

We have examined how the results depend on the definition of the exact observable. If each pair of dipoles is allowed to interact inelastically several times, with a Poisson distribution with mean $2\gamma$, as the AGK rules imply for the eikonal $S$-matrix $e^{-\gamma}$ of a pair, the cross section $\sigma_0$ is unchanged and the mean is multiplied by $2\gamma/p$, which is $1.03$ for $\gamma=0.025$ and $1.10$ for $\gamma=0.1$; in the rest frame of the target the entropy changes by $0.03$ at $\tilde Y=400$ in the first case and by $0.13$ at $\tilde Y=200$ in the second. None of the conclusions below depends on this choice. We have also verified that the exact distribution satisfies the evolution equation for the cross sections of fixed multiplicity (Sec.~\ref{sec:backward}).

\begin{figure*}[tb]
\includegraphics[width=0.95\textwidth]{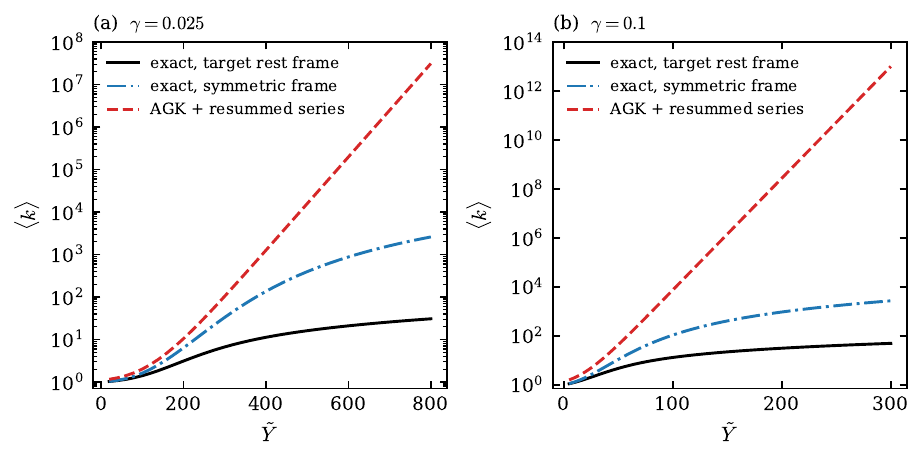}
\caption{\label{fig:utmmean} Mean number of inelastic dipole-dipole interactions in the UTM as a function of rapidity for (a) $\gamma=0.025$ and (b) $\gamma=0.1$. Solid and dash-dotted lines: exact result of \eq{eq:cutsutm} in the rest frame of the target and in the symmetric frame. Dashed lines: AGK cutting rules applied to the resummed series, \eq{eq:meanAGK}.}
\end{figure*}

\subsubsection{Mean multiplicity}

Figure~\ref{fig:utmmean} shows the mean of the three distributions. They agree at small $\tilde Y$, where $\langle n\rangle\ll1/\gamma$ and the model is close to the BFKL cascade, and separate at larger $\tilde Y$. The exact mean in the rest frame is $\langle k\rangle = p\,\langle n\rangle/\sigma_{in}$, where $\sigma_{in}=\sum_{k\geq1}\sigma_k$; since the rate of emission in \eq{eq:utm} tends to unity at $n\gg1/\gamma$, $\langle n\rangle$ grows linearly with $\tilde Y$ at large rapidity. In the symmetric frame $\langle k\rangle\simeq p\,\langle n\rangle^2_{\tilde Y/2}$ grows quadratically. The AGK mean of \eq{eq:meanAGK} grows exponentially. For $\gamma=0.025$ and $\tilde Y=400$, where the resummed series reproduces the exact $S$-matrix within $2.2\%$, the three values are $11.3$, $136$ and $1.26\times10^{3}$; for $\gamma=0.1$ and $\tilde Y=200$ they are $31.2$, $940$ and $2.7\times10^{8}$. Of the two frames, the symmetric one is closer to the AGK result; the region in which they agree is determined in Sec.~\ref{sec:origin}.

\subsubsection{Absence of KNO scaling}

\begin{figure*}[tb]
\includegraphics[width=0.95\textwidth]{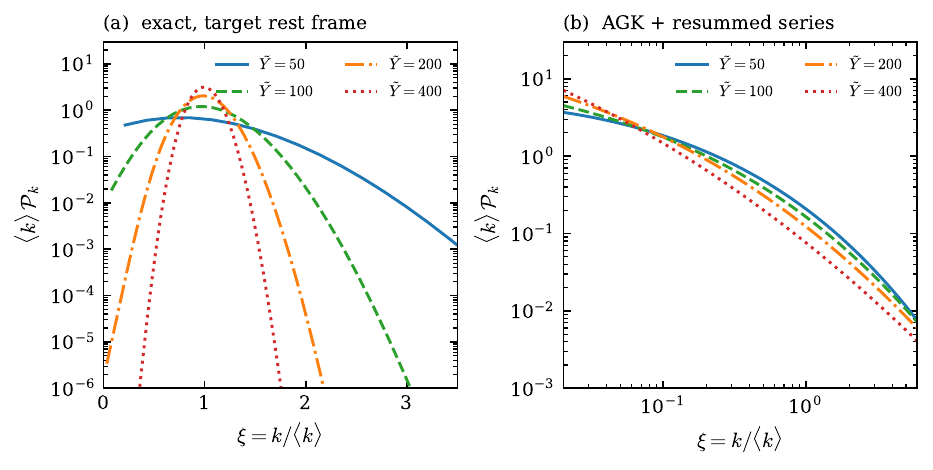}
\caption{\label{fig:utmkno} Multiplicity distributions in the UTM for $\gamma=0.1$ in the KNO variables. (a) Exact distribution of \eq{eq:cutsutm} in the rest frame of the target. (b) AGK cutting rules applied to the resummed series, \eq{eq:sAGKutm}. Note the different scales of the horizontal axes.}
\end{figure*}

Figure~\ref{fig:utmkno} shows the distributions as functions of $\xi=k/\langle k\rangle$. The exact distribution in the rest frame of the target becomes narrower as $\tilde Y$ increases: the ratio $\langle k^2\rangle/\langle k\rangle^2$ is $1.32$, $1.11$, $1.04$ and $1.02$ at $\tilde Y=50$, $100$, $200$ and $300$ for $\gamma=0.1$. At large $\tilde Y$ the variance approaches the mean, as for a Poisson distribution, which is what \eq{eq:utm} gives when the emission rate is saturated. The distribution of \eq{eq:sAGKutm} remains broad, with a slowly varying shape which extends over two orders of magnitude in $\xi$. The exact distribution does not have KNO scaling.

\subsubsection{Entropy}

\begin{figure}[tb]
\includegraphics[width=0.95\columnwidth]{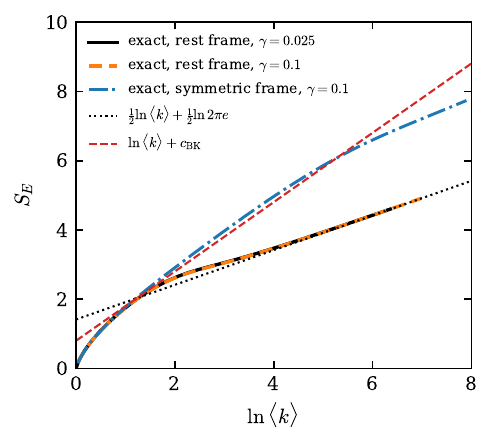}
\caption{\label{fig:utment} Entropy of the exact distribution of \eq{eq:cutsutm} in the UTM as a function of $\ln\langle k\rangle$, in the rest frame of the target for $\gamma = 0.025$ (solid line) and $\gamma=0.1$ (long-dashed line, which coincides with the solid one), and in the symmetric frame for $\gamma=0.1$ (dash-dotted line). Each curve ends at the largest rapidity computed: $\tilde Y=1.2\times10^{4}$ and $6\times10^{3}$ in the rest frame, where $\langle k\rangle$ grows only linearly with $\tilde Y$, and $\tilde Y=300$ in the symmetric frame. Dotted line: entropy of a Poisson distribution, \eq{eq:Spoisson}. Short-dashed line: $\ln\langle k\rangle + c_{\rm BK}$, shown for reference. }
\end{figure}

The entropy of the exact distribution is shown in Fig.~\ref{fig:utment}. In the rest frame of the target it approaches\verificado
\beq \label{eq:Spoisson}
S_E \,=\, \h \ln\langle k\rangle + \h\ln\Lb 2\pi e\Rb ,
\eeq
the entropy of a Poisson distribution with large mean: it grows as one half of the logarithm of the mean multiplicity, and the curves for the two values of $\gamma$ coincide. Since in this frame $\langle k\rangle$ grows only linearly with rapidity, large values of $\ln\langle k\rangle$ require very large $\tilde Y$; at $\tilde Y=6\times10^{3}$ for $\gamma=0.1$ we find $\langle k\rangle=1.08\times10^{3}$ and $S_E=4.915$, while \eq{eq:Spoisson} gives $4.913$. In the symmetric frame the entropy is close to $\ln\langle k\rangle + {\rm const}$ at intermediate rapidities, with $S_E-\ln\langle k\rangle = 0.94$, $0.98$ and $0.91$ at $\tilde Y=50$, $75$ and $100$ ($\gamma=0.1$), and grows more slowly at larger $\tilde Y$: the same difference is $0.60$ at $\tilde Y=150$, $0.27$ at $\tilde Y=200$ and $-0.13$ at $\tilde Y=290$, where the distribution is narrow, $\langle k^2\rangle/\langle k\rangle^2-1 = 0.05$. At asymptotically large rapidities, where $k\simeq p\,n\,m$ with $n$ and $m$ distributed as in the rest frame, the variance of $k$ grows as $\tilde Y^3\propto\langle k\rangle^{3/2}$, and the entropy approaches $\frac{3}{4}\ln\langle k\rangle+{\rm const}$: in the Gaussian approximation $S_E-\frac{3}{4}\ln\langle k\rangle = 1.83$, $1.51$, $1.39$ and $1.35$ at $\tilde Y=290$, $10^{3}$, $4\times10^{3}$ and $1.6\times10^{4}$ for $\gamma=0.1$.

\subsubsection{Frame dependence}

The comparison of the two frames shows that the number of inelastic dipole-dipole interactions is not a frame-independent quantity in the UTM, although the $S$-matrix is. The mean differs by a factor of $12$ at $\tilde Y=400$ for $\gamma=0.025$ and by a factor of $30$ at $\tilde Y=200$ for $\gamma=0.1$, and the shape and the entropy of the distribution are different as well. The AGK distribution of \eq{eq:sAGKutm} is, by construction, a single function of $\tilde Y$. It overestimates the exact mean in both frames, by a factor which increases exponentially with $\tilde Y$.

To summarize, in a model in which the resummed multi-Pomeron series reproduces the exact scattering amplitude, the multiplicity distribution derived from the same series with the AGK cutting rules does not reproduce the exact distribution of the inelastic interactions at large rapidity, neither in the mean, nor in the shape, nor in the entropy. At which rapidity the two start to differ, and why, is the subject of the next section.

\section{Origin of the discrepancy}
\label{sec:origin}

In this section we show where the AGK distribution of \eq{eq:sAGKutm} agrees with the exact one, why it fails outside this region, and why the same series nevertheless reproduces the scattering amplitude at all rapidities.

\subsection{Probabilistic meaning of the AGK distribution}
\label{sec:meaning}

The integrand of \eq{eq:sAGKutm} can be written as\verificado
\beq \label{eq:identity}
\frac{\Lb 2\tau\gamma N_j\Rb^n}{\Lb 1+2\tau\gamma N_j\Rb^{n+1}} \,=\, \intl^\infty_0 dx\, e^{-x}\;\frac{\Lb 2\gamma N_j\,\tau x\Rb^n}{n!}\,e^{-2\gamma N_j \tau x} ,
\eeq
so that
\beq \label{eq:twocascades}
\sigma^{\rm AGK}_n = \Big\langle \intl^\infty_0 \!d\tau\, e^{-\tau}\!\intl^\infty_0 \!dx\, e^{-x}\,\frac{\Lb 2\gamma\, n_1 n_2\Rb^n}{n!}\,e^{-2\gamma\, n_1 n_2}\Big\rangle_{j} ,
\eeq
with $n_1 n_2 = N_j\,\tau\,x$, where $\langle\dots\rangle_j$ denotes the average over the Poisson distribution of $j$ with mean $\tilde Y$. Equation~(\ref{eq:twocascades}) has a direct interpretation. The numbers of dipoles $n_1$ and $n_2$ in the two colliding systems have independent exponential distributions, with the product of the mean values equal to $N_j$; each of the $n_1 n_2$ pairs of dipoles interacts inelastically a number of times which has a Poisson distribution with mean $2\gamma$, as the AGK rules prescribe for the eikonal $S$-matrix $e^{-\gamma}$ of a pair. The exponential distribution is the large-multiplicity limit of the geometric distribution of the number of dipoles in the BFKL cascade~\cite{Mueller:1993rr,Mueller:1996te,Kharzeev:2017qzs}, i.e., in a cascade without saturation. The average over $j$ originates from the expansion of $\exp\Lb\Delta_k\tilde Y\Rb$ used in the resummation; it gives $\langle N_j\rangle_j = \exp\Lb\Lb e^{\gamma}-1\Rb\tilde Y\Rb$, the single-Pomeron exchange.

The AGK cutting rules applied to the resummed series thus give the distribution of the number of inelastic interactions between two \emph{dilute} cascades. The same is true for \eq{eq:Sutm}: it is the $S$-matrix $\langle e^{-\gamma n_1 n_2}\rangle$ for the same two cascades. In QCD, \eq{eq:sdd} has the structure of \eq{eq:twocascades}, with the Gaussian integral over $\lambda$ in place of the average over $j$, and \eq{eq:sBK} without the factor $1/n$ is the analogous expression for a single cascade (see Sec.~\ref{sec:mixture}).

This interpretation tells us where to expect agreement with the exact distribution of \eq{eq:cutsutm}: in a frame in which both cascades are dilute,
\beq \label{eq:window}
\gamma\,\langle n\rangle_{\tilde Y_0} \ll 1, \qquad \gamma\,\langle n\rangle_{\tilde Y-\tilde Y_0} \ll 1 .
\eeq
The widest range of rapidities is obtained in the symmetric frame, where \eq{eq:window} holds for $\tilde Y < 2\,\tilde Y_s$ with
\beq \label{eq:Ys}
\tilde Y_s \,\equiv\, \frac{\ln\Lb 1/\gamma\Rb}{e^{\gamma}-1} ,
\eeq
the rapidity at which the single-Pomeron exchange becomes of the order of unity. In the rest frame of the target the corresponding range is $\tilde Y<\tilde Y_s$. For $\tilde Y_s<\tilde Y<2\tilde Y_s$ the mean number of inelastic interactions in the symmetric frame, $2\gamma\langle n\rangle^2_{\tilde Y/2}$, is large (up to values of the order of $1/\gamma$) and the $S$-matrix is small, while each of the cascades is still dilute. This is the kinematic window of the approach of Refs.~\cite{Mueller:1996te,Salam:1995uy,Iancu:2003zr}, on which \eq{eq:MPSI} is based.

\begin{table*}[tb]
\caption{\label{tab:sym} Number of inelastic dipole-dipole interactions in the UTM in the symmetric frame for $\gamma=0.005$ ($\tilde Y_s=1057$). Exact: \eq{eq:cutsutm}. Two dilute cascades: \eq{eq:twocascades} without the average over $j$ and with $N_j$ replaced by the exact $\langle n\rangle^2_{\tilde Y/2}$. Last column: mean of the AGK distribution of \eq{eq:sAGKutm}, \eq{eq:meanAGK}.}
\begin{ruledtabular}
\begin{tabular}{rcrccrccr}
 & & \multicolumn{3}{c}{exact} & \multicolumn{3}{c}{two dilute cascades, exact $\langle n\rangle$} & AGK, resummed series\\
$\tilde Y$ & $\gamma\langle n\rangle_{\tilde Y/2}$ & $\langle k\rangle$ & $\langle k^2\rangle/\langle k\rangle^2$ & $S_E$ & $\langle k\rangle$ & $\langle k^2\rangle/\langle k\rangle^2$ & $S_E$ & $\langle k\rangle$\\
\hline
600 & 0.022 & 1.27 & 1.25 & 0.65 & 1.35 & 1.33 & 0.76 & 1.36\\
1200 & 0.093 & 5.06 & 2.46 & 2.47 & 5.40 & 2.74 & 2.52 & 6.15\\
1800 & 0.34 & 47.0 & 2.59 & 4.79 & 48.8 & 3.73 & 4.73 & 87.1\\
2400 & 0.93 & 347 & 1.93 & 6.84 & 353 & 3.94 & 6.69 & $1.68\times10^{3}$\\
3000 & 1.9 & $1.50\times10^{3}$ & 1.48 & 8.23 & $1.51\times10^{3}$ & 3.98 & 8.14 & $3.40\times10^{4}$\\
4000 & 4.2 & $6.94\times10^{3}$ & 1.17 & 9.38 & $6.98\times10^{3}$ & 4.00 & 9.66 & $5.10\times10^{6}$\\
\end{tabular}
\end{ruledtabular}
\end{table*}

Table~\ref{tab:sym} tests this expectation for $\gamma=0.005$, for which the scales $1$, $1/\gamma$ and $\tilde Y_s$ are well separated. Three regions can be distinguished.

For $\gamma\langle n\rangle_{\tilde Y/2}\lesssim 0.1$ the AGK distribution is close to the exact one. The means differ by $7\%$ at $\tilde Y=600$ and by $22\%$ at $\tilde Y=1200$, where $\langle k\rangle\approx5$. The AGK distribution of \eq{eq:sAGKutm} has $\langle k^2\rangle/\langle k\rangle^2=1.34$ and $2.91$ and $S_E=0.78$ and $2.65$ at these two rapidities, to be compared with the exact values $1.25$ and $2.46$, and $0.65$ and $2.47$. At the smaller rapidity, where $\langle n\rangle_{\tilde Y/2}=4.4$, a part of the difference is due to the discreteness of the number of dipoles, which the exponential distribution neglects. In this region the entropy of the exact distribution approaches $S_E = \ln\langle k\rangle + c$, with $S_E-\ln\langle k\rangle=0.85$ at $\tilde Y=1200$; for two dilute cascades at large mean multiplicity the KNO function is $2K_0\Lb 2\sqrt{\xi}\Rb$ and $c = 0.81$.

For $0.1\lesssim\gamma\langle n\rangle_{\tilde Y/2}\lesssim 1$ the mean of the AGK distribution departs from the exact one, by a factor of $1.9$ at $\tilde Y=1800$ and $4.8$ at $\tilde Y=2400$. The reason is that the exact $\langle n\rangle_{\tilde Y/2}$ falls below the single-Pomeron exchange (Sec.~\ref{sec:multiP}). If the exact $\langle n\rangle$ is used in the distribution of two dilute cascades, the mean is reproduced by construction, and the entropy is reproduced within $0.15$, but the distribution is broader than the exact one.

For $\gamma\langle n\rangle_{\tilde Y/2}\gtrsim1$ both cascades are saturated. The exact distribution becomes narrow, $\langle k^2\rangle/\langle k\rangle^2\to1$, while for two dilute cascades this ratio is $4$, and the mean of the AGK distribution exceeds the exact one by a factor which grows exponentially: $23$ at $\tilde Y=3000$ and $7.3\times10^{2}$ at $\tilde Y=4000$.

\begin{figure*}[tb]
\includegraphics[width=0.95\textwidth]{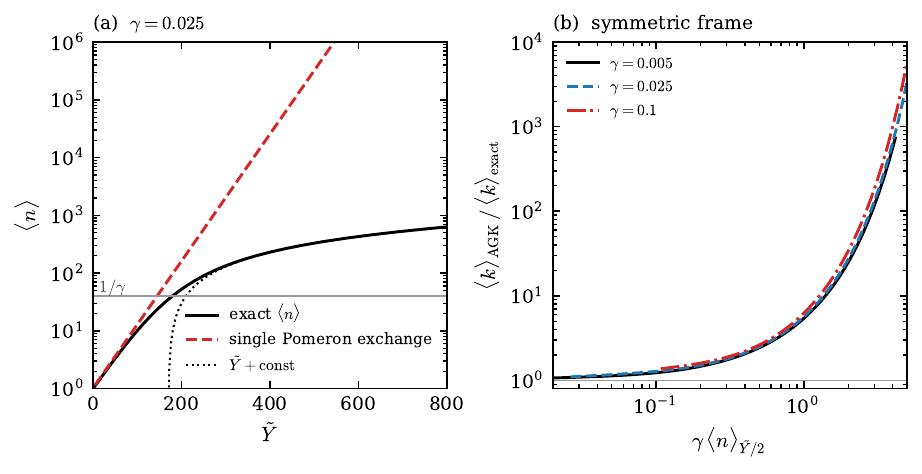}
\caption{\label{fig:origin} (a) Mean number of dipoles in the UTM for $\gamma=0.025$: exact solution of \eq{eq:utm} (solid line), the single-Pomeron term of \eq{eq:nseries} (dashed line), and the linear growth at large rapidity (dotted line). The horizontal line marks $\langle n\rangle=1/\gamma$. (b) Ratio of the mean of the AGK distribution, \eq{eq:meanAGK}, to the exact mean number of inelastic interactions in the symmetric frame, as a function of $\gamma\langle n\rangle_{\tilde Y/2}$, for three values of $\gamma$.}
\end{figure*}

Figure~\ref{fig:origin}(b) shows that the ratio of the two means in the symmetric frame is, to a good accuracy, a function of the single variable $\gamma\langle n\rangle_{\tilde Y/2}$ for the three values of $\gamma$ considered: it is $1.2$--$1.4$ at $\gamma\langle n\rangle_{\tilde Y/2}=0.1$, $1.7$--$1.9$ at $0.3$, and $5$--$6$ at $1$. At the upper end of the window, $\tilde Y=2\tilde Y_s$, one has $\gamma\langle n\rangle_{\tilde Y/2}\approx0.6$ and the ratio is about $3$. The corrections to the AGK distribution are therefore not exponentially small inside the window: the relative difference of the means is $0.2$--$0.4$ already at $\gamma\langle n\rangle_{\tilde Y/2}=0.1$ and grows faster than linearly with this variable. In the rest frame of the target the ratio of the means is a function of $\gamma\langle n\rangle_{\tilde Y}$, equal to $1.2$--$1.4$, $1.5$--$1.6$ and $2.6$--$3$ at $\gamma\langle n\rangle_{\tilde Y}=0.1$, $0.3$ and $1$.

\subsection{Multi-Pomeron contributions to the mean multiplicity}
\label{sec:multiP}

The mean of the exact distribution of \eq{eq:cutsutm} is
\beq \label{eq:meanexact}
\langle k\rangle\,\sigma_{in} \,=\, p\;\langle n\rangle_{\tilde Y_0}\,\langle n\rangle_{\tilde Y-\tilde Y_0} ,
\eeq
and the mean number of dipoles follows from \eq{eq:Zutm}:\verificado
\bea \label{eq:nseries}
\langle n\rangle_{\tilde Y} &=& \frac{\partial Z}{\partial u}\bigg|_{u=1} = \sum_{k=1}^{\infty} C_k\,\Phi'_k\Lb 1\Rb\,e^{\Delta_k \tilde Y} ,\nn\\
\Phi'_k\Lb 1\Rb &=& \prod_{l=1}^{k-1}\Lb e^{-\gamma l}-1\Rb \,\simeq\, \Lb-\gamma\Rb^{k-1}\Lb k-1\Rb! .
\eea
The term with $k=1$ is the single-Pomeron exchange, $e^{\gamma}\exp\Lb\Delta_1\tilde Y\Rb$. Keeping only this term for both cascades in \eq{eq:meanexact} gives $p\,e^{2\gamma}\exp\Lb\Delta_1\tilde Y\Rb\simeq 2\gamma\exp\Lb\Lb e^{\gamma}-1\Rb\tilde Y\Rb$, which does not depend on $\tilde Y_0$ and coincides with the numerator of \eq{eq:meanAGK}. The mean of the AGK distribution is thus the single-Pomeron part of the exact mean.

The terms with $k\geq2$ in \eq{eq:nseries} do not vanish. They alternate in sign, and for $\gamma k\ll1$ the term with $k$ Pomerons is of the relative order of $\Lb k-1\Rb!\,\Lb\gamma\,e^{\Delta_1\tilde Y}\Rb^{k-1}$. Equation~(\ref{eq:nseries}) is an asymptotic series; at small $\tilde Y$, truncated at its smallest term, it reproduces the solution of \eq{eq:utm} (for $\gamma=0.025$ and $\tilde Y=20$ both give $\langle n\rangle=1.6328$, while the single-Pomeron term is $1.7011$). Table~\ref{tab:nmean} and Fig.~\ref{fig:origin}(a) show that the contribution of the multi-Pomeron exchanges increases with rapidity and at $\langle n\rangle\gtrsim1/\gamma$ cancels the exponential growth of the single-Pomeron term, leaving the linear growth $\langle n\rangle\simeq\tilde Y+{\rm const}$.

\begin{table}[tb]
\caption{\label{tab:nmean} Mean number of dipoles in the UTM for $\gamma=0.025$: exact solution of \eq{eq:utm} and the single-Pomeron term of \eq{eq:nseries}.}
\begin{ruledtabular}
\begin{tabular}{rrrr}
$\tilde Y$ & exact $\langle n\rangle$ & single Pomeron & ratio\\
\hline
20 & 1.63 & 1.70 & 1.04\\
80 & 6.60 & 7.77 & 1.18\\
160 & 30.0 & 58.9 & 1.96\\
240 & 81.8 & 446 & 5.45\\
400 & 231 & $2.56\times10^{4}$ & 111\\
\end{tabular}
\end{ruledtabular}
\end{table}

This is to be compared with the AGK cancellation, according to which only the single-Pomeron exchange contributes to the mean number of cut Pomerons. In the rest frame of the target the generating function of the exact distribution is obtained from $Z\Lb u,\tilde Y\Rb$ by the replacement $u\to 1-p\Lb 1-x\Rb$ [see \eq{eq:genrest} of Appendix~\ref{app:exact}], so that the cuts are counted by the variable $t=1-u$. The rules of \eq{eq:AGK} for the term with $k$ Pomerons are equivalent to the assumption that this term is proportional to $t^{k}$. The exact eigenfunctions do not have this property:\verificado
\beq \label{eq:Philinear}
\Phi_k\Lb 1-t\Rb \,=\, -\,t\prod_{l=1}^{k-1}\Lb e^{-\gamma l}-1\Rb\,+\,\mathcal{O}\Lb t^2\Rb ,
\eeq
i.e., they vanish linearly for all $k$. For small $\gamma$ and $u=e^{-s}$ one has $\Phi_k\Lb e^{-s}\Rb\simeq\Lb-\gamma\Rb^k\,\Gamma\Lb k+s/\gamma\Rb/\Gamma\Lb s/\gamma\Rb$, which equals $\Lb-\gamma\Rb^k k!$ at $s=\gamma$, the value used in Refs.~\cite{Kovner:2024tin,Levin:2024wtl}. The amplitude requires $\Phi_k$ at this single point, and the approximation is sufficient for it. The multiplicity distribution requires $\Phi_k$ as a function of $u$, and replacing it by $k!\,\Lb -s\Rb^{k}$ is what makes the AGK rules of \eq{eq:AGK} applicable term by term. We note the similarity of the ratio $\Gamma\Lb k+s/\gamma\Rb/\Gamma\Lb s/\gamma\Rb$ with the factor $\Gamma\Lb\omega+n\Rb/\Gamma\Lb\omega\Rb$ of \eq{eq:old}.

In physical terms, the terms with $k\geq2$ in \eq{eq:nseries} describe the saturation of the emission rate in \eq{eq:utm}, i.e., the interaction of the Pomerons inside the wave function of one dipole. They are absent in the BFKL cascade, for which $\langle n\rangle$ is given by the single-Pomeron exchange at all rapidities.

\subsection{Why the amplitude is reproduced and the multiplicity is not}
\label{sec:rare}

\begin{figure}[tb]
\includegraphics[width=0.95\columnwidth]{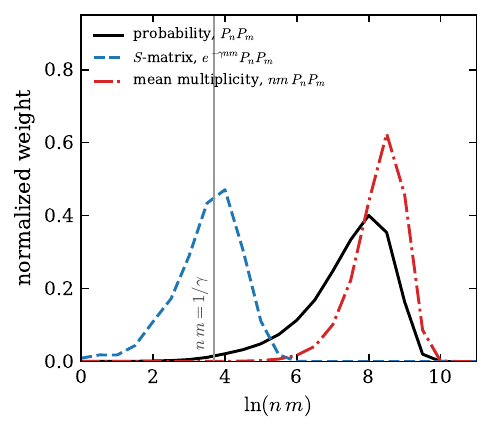}
\caption{\label{fig:configs} Distribution in $\ln\Lb n\,m\Rb$ of the configurations of the two cascades in the symmetric frame for $\gamma=0.025$ and $\tilde Y=400$, weighted with the probability (solid line), with the contribution to the $S$-matrix (dashed line), and with the contribution to the mean number of inelastic interactions (dash-dotted line). Each curve is normalized to unity. The vertical line marks $n\,m=1/\gamma$.}
\end{figure}

The $S$-matrix of \eq{eq:Sframe} and the mean of \eq{eq:meanexact} are sums over the same configurations of the two cascades, but with different weights: $e^{-\gamma n m}$ and $n\,m$. Figure~\ref{fig:configs} shows the three distributions in the symmetric frame for $\gamma=0.025$ and $\tilde Y=400$, where $\langle n\rangle_{\tilde Y/2}=52.6$, above $1/\gamma=40$. The typical configurations have $n\,m\sim\langle n\rangle^2_{\tilde Y/2}=2.8\times10^{3}$, and the configurations which determine the multiplicity have still larger $n\,m$. The $S$-matrix, on the contrary, is determined by the configurations with $n\,m\sim1/\gamma$: the mean value of $n\,m$ with the weight $e^{-\gamma nm}$ is $48$. These are rare configurations (the probability to have $n\,m\leq200$ is $6.5\%$), in agreement with the general analysis of Ref.~\cite{Iancu:2003zr}. They are asymmetric: with the same weight, the mean value of the smaller of the two numbers is $2.4$ and that of the larger one is $25$. One of the cascades is almost empty, and the other contains about half of its mean number of dipoles.

For configurations with a small number of dipoles the dilute description is accurate at any rapidity. Equation~(\ref{eq:utm}) is triangular: $P_n\Lb\tilde Y\Rb$ is determined by the emission rates $1-e^{-\gamma m}$ with $m\leq n$ only. For $n\ll1/\gamma$ these rates are those of the BFKL cascade, up to corrections of the relative order of $\gamma n$, whatever the mean number of dipoles is. The probabilities of the configurations with few dipoles are therefore not affected by the saturation of the typical ones.

This gives a qualitative explanation of both observations. The resummed series is the $S$-matrix of two dilute cascades (Sec.~\ref{sec:meaning}). It reproduces the exact $S$-matrix at all rapidities because, in the symmetric frame, the exact $S$-matrix is built from configurations with far fewer dipoles than the typical ones, in which the saturation effects are weak or moderate. The multiplicity distribution is built from the typical configurations, which at $\tilde Y>2\tilde Y_s$ are saturated. The agreement of the amplitudes tests the probabilities $P_n$ at small $n$ and gives no information on the region of $n$ that determines the multiplicities.

Two remarks are in order. First, the separation is not sharp: for the parameters of Fig.~\ref{fig:configs}, $20\%$ of the $S$-matrix comes from configurations in which the larger cascade has more than $1/\gamma$ dipoles, and the accuracy of $2\%$ of \eq{eq:Sutm} relies on the average over $j$, which goes beyond the picture of two BFKL cascades with fixed mean multiplicities. Second, in the rest frame of the target the same $S$-matrix is built from configurations with $n\approx\langle n\rangle/2\gg1/\gamma$ ($117$ versus $\langle n\rangle = 231$ for the same parameters); the equality of the two representations is the $t$-channel unitarity of the model.

\subsection{Evolution equations for the cross sections of fixed multiplicity}
\label{sec:backward}

Reference~\cite{Levin:2024wtl} shows that the AGK cutting rules are equivalent to evolution equations for $\sigma_n$ in the parton cascade, in which the derivative of the cross section for dipole-dipole scattering is expressed through the cross section for the scattering of a dipole with two dipoles, $d\sigma^{d+d}_n/d\tilde Y\propto\sigma^{d+2d}_n-\sigma^{d+d}_n$. It is therefore necessary to check whether the exact distribution of \eq{eq:cutsutm} is consistent with these equations.

Let $Z^{(m)}\Lb u,\tilde Y\Rb$ be the generating function of \eq{eq:utm} with the initial condition $P_n\Lb0\Rb=\delta_{nm}$. Considering the first step of the evolution one obtains the backward equation
\beq \label{eq:backward}
\frac{\partial Z^{(1)}}{\partial\tilde Y} \,=\,\Lb 1-e^{-\gamma}\Rb\Lb Z^{(2)}\Lb u,\tilde Y\Rb - Z^{(1)}\Lb u,\tilde Y\Rb\Rb ,
\eeq
valid for any $u$. With $u=1-p\Lb1-x\Rb$, the coefficient of $x^n$ in \eq{eq:backward} is an equation of the form of Ref.~\cite{Levin:2024wtl}, with $\sigma_n^{d+2d}$ defined as the cross section for a projectile which starts as two dipoles. The exact distribution in the rest frame of the target satisfies it identically. We have verified this numerically by solving \eq{eq:utm} with the two initial conditions (with the Poisson form of the cuts, $u=e^{-2\gamma(1-x)}$): for $\gamma=0.1$ and $n=1$, $3$ and $10$ the two sides of the equation agree to a relative accuracy of $10^{-5}$ for $5\leq\tilde Y\leq55$.

For the BFKL cascade the dipoles evolve independently, $Z^{(2)}=\Lb Z^{(1)}\Rb^2$, and \eq{eq:backward} becomes the nonlinear equation of Refs.~\cite{Mueller:1993rr,Kovchegov:1999ji,Levin:2024wtl}. In the UTM two dipoles do not evolve independently, since the emission rate for $n=2$ is $1-e^{-2\gamma}\neq2\Lb1-e^{-\gamma}\Rb$, and $Z^{(2)}\neq\Lb Z^{(1)}\Rb^2$. The difference between \eq{eq:cutsutm} and \eq{eq:sAGKutm} is therefore not in the form of the evolution equation but in the input for $\sigma^{d+2d}_n$: the exact one follows from $Z^{(2)}$ at any $\gamma n$, while in Ref.~\cite{Levin:2024wtl} the equation is written for the Borel images of the cross sections in the region of small $\gamma$.

\section{Discussion: implications for QCD}
\label{sec:discussion}

\subsection{What the KNO functions of QCD describe}

The results of Secs.~\ref{sec:qcd} and \ref{sec:origin} can be put together. The KNO functions which follow from the Gaussian representation are mixtures of the distributions of dilute cascades: one cascade for the BK amplitude, \eq{eq:mixBK}, and two cascades for dipole-dipole scattering, \eq{eq:mixdd}. In the UTM the same procedure gives \eq{eq:twocascades}, in which the Gaussian weight is replaced by the Poisson distribution of $j$. In that model we know the exact answer: the distribution of two dilute cascades is correct as long as the cascades which determine the multiplicity are dilute, \eq{eq:window}, and fails when they saturate, although the amplitude is reproduced in both regimes.

The question for QCD is therefore whether the dipole configurations that determine the multiplicity are dilute in the kinematic region where the approach is applied. The coefficients $C_n$ of \eq{eq:Cn} are extracted from the solution of the BK equation deep in the saturation region, $z\gg1$, and the KNO functions are derived in the same region. In the UTM the analogous region is $\gamma\langle n\rangle\gg1$, where the dilute description of the multiplicity fails. For dipole-dipole scattering the window of \eq{eq:window} suggests that the range of validity is $z\lesssim 2 z_s$, where $z_s$ is the value of $z$ at which the single-Pomeron exchange is of the order of unity; this is the known range of validity of the approach based on \eq{eq:MPSI}~\cite{Mueller:1996te,Iancu:2003zr}. For the BK amplitude, in which the target is dense from the beginning, the UTM has no window of this kind.

We stress the limits of this argument. The models considered here have no transverse dimensions. In QCD the saturation of the amplitude for a dipole of a given size does not stop the growth of the number of gluons, which continues at larger transverse momenta, and for dipole-nucleus scattering the inclusive gluon production has been calculated directly in QCD~\cite{Kovchegov:2001sc}, without reference to the multi-Pomeron series. Our results do not show that the AGK cutting rules are violated in QCD. They show that the agreement of a resummed multi-Pomeron series with the exact amplitude, which is the test passed in the UTM and invoked in Ref.~\cite{Levin:2025cgd}, is not sufficient to validate the multiplicity distribution derived from the same series: the two quantities are sensitive to different configurations of the cascades (Sec.~\ref{sec:rare}).

\subsection{Nonuniqueness of the sum: an indeterminate moment problem}

The mixture representation of Sec.~\ref{sec:mixture} makes precise the statement of Ref.~\cite{Levin:2025cgd} that different summation procedures can lead to different results. The Gaussian representation replaces the factor $\exp\Lb n^2/2\Rb$ by the $n$-th moment of a log-normal distribution, $\exp\Lb n^2/2\Rb = \int d\mu\Lb\lambda\Rb\,e^{n\lambda}$ with $d\mu = e^{-\lambda^2/2}d\lambda/\sqrt{2\pi}$. Only integer values of $n$ enter the series, and the log-normal distribution is the classical example of a distribution which is not determined by its integer moments~\cite{Heyde:1963}. For instance, all the measures
\beq \label{eq:family}
d\mu_{\epsilon}\Lb\lambda\Rb \,=\,\Lb 1+\epsilon\,\sin\Lb 2\pi\lambda\Rb\Rb\, e^{-\h\lambda^2}\,\frac{d\lambda}{\sqrt{2\pi}}, \qquad |\epsilon|\leq1,
\eeq
are positive and have the same moments $\exp\Lb n^2/2\Rb$ for all integer $n$, since $\int d\mu\,e^{n\lambda}\sin\Lb2\pi\lambda\Rb\propto{\rm Im}\,e^{\Lb n+2\pi i\Rb^2/2}=0$. Each of them reproduces every term of the series of \eq{eq:fan}, and each leads to a positive KNO function,
\beq \label{eq:Psieps}
\Psi_{\epsilon}\Lb\xi\Rb \,=\,\intl d\mu_{\epsilon}\Lb\lambda\Rb\; e^{\lambda}\, e^{-\xi\, e^{\lambda}} .
\eeq
The choice $\epsilon=0$ of Ref.~\cite{Levin:2025cgd} is the simplest one, but it is not singled out by the series.

The numerical size of this ambiguity is small where most of the probability is concentrated and increases in the tail. At $\xi\gtrsim10^{2}$ a steepest-descent estimate with the saddle point of \eq{eq:spBK} gives $\left|\Psi_{\epsilon}-\Psi_{\rm BK}\right|/\Psi_{\rm BK}\simeq|\epsilon|\exp\Lb-2\pi^2/\Lb1+W\Rb\Rb$, with $W=W\Lb e\xi\Rb$. Numerically, for $\epsilon=1$ and an arbitrary phase of the oscillating term, the relative difference is $5\times10^{-4}$ at $\xi=1$, $0.7\%$ at $\xi=10$, $3\%$ at $\xi=10^{2}$, and $12\%$ at $\xi=10^{4}$. The leading log-normal behavior of \eq{eq:tailBK} is common to the whole family. Within this family the ambiguity is therefore of little practical importance at $\xi\lesssim10$; we have not found a criterion that would restrict the admissible measures to this family. The contour representation of \eq{eq:old}, which organizes the same large-$n$ behavior in a different way, leads to a function that is not positive (Sec.~\ref{sec:assumptions}).

\subsection{Entropy}

The relation $S_E=\ln\bar n$ follows from KNO scaling alone, \eq{eq:entropy}, and identifies the entropy with $\ln\Lb xG\Rb$ only to the extent that the mean multiplicity is given by the single-Pomeron exchange. We have found three qualifications. (a) In QCD the constant terms are $c_{\rm BK}=0.81$ and $c_{\rm dd}=0.09$, \eq{eq:consts}, and the scaling itself requires that the coefficients $C_n$ have no power-like factors at small $n$ (Sec.~\ref{sec:assumptions}). (b) In the UTM, in the symmetric frame, the entropy of the exact distribution follows $S_E\simeq\ln\langle k\rangle+c$ with $c$ between $0.85$ and $1.0$ for $0.1\lesssim\gamma\langle n\rangle_{\tilde Y/2}\lesssim2$ (Table~\ref{tab:sym}, $\gamma=0.005$), close to the value $c=0.81$ for two dilute cascades; however, $\langle k\rangle$ in this relation is the exact mean, which in most of this range is below the single-Pomeron exchange. (c) When the cascades are saturated the coefficient of the logarithm is smaller than unity: the entropy approaches $\h\ln\langle k\rangle$ plus a constant in the rest frame of the target, \eq{eq:Spoisson}, already at $\tilde Y>\tilde Y_s$, and $\frac{3}{4}\ln\langle k\rangle$ plus a constant in the symmetric frame at asymptotically large rapidities (Sec.~\ref{sec:utm}). In the UTM the relation between the entropy and the mean multiplicity thus depends on the regime and on the frame.

\subsection{Limitations}

Several limitations of this work should be kept in mind. The exact tests are performed in zero transverse dimensions, where the multiplicity is represented by the number of inelastic dipole-dipole interactions; these models contain no produced gluons, and the convolution of \eq{eq:conv} is not tested. The number of inelastic interactions is defined in a given frame, and we have found that it is frame dependent (Sec.~\ref{sec:utm}); a frame-independent definition of the multiplicity in these models would require an explicit treatment of the final state, which is beyond the scope of this paper. In QCD the coefficients $C_n$ at small $n$ are not known, and we have not attempted to determine them: the statements of Sec.~\ref{sec:assumptions} concern the consequences of the assumptions made about them. Finally, the results of Secs.~\ref{sec:qcd} and \ref{sec:assumptions} are properties of the integral representations of Refs.~\cite{Levin:2024ire,Levin:2025cgd} and do not depend on the toy models.

\section{Conclusions}
\label{sec:conclusions}

We have reexamined the multiplicity distributions for the BK amplitude and for dipole-dipole scattering which follow from the sum of large Pomeron loops and the AGK cutting rules~\cite{Levin:2024ire,Levin:2025cgd}, and we have tested the approach in two exactly solvable models. Our conclusions are the following.

(1) The KNO functions obtained with the Gaussian representation are given in closed form, at the steepest-descent accuracy, by Eqs.~(\ref{eq:PsiBKsd}) and (\ref{eq:Psiddsd}) in terms of the Lambert function. Their tails are log-normal, $\ln\Psi_{\rm BK}\simeq-\h\ln^2\xi$ and $\ln\Psi_{\rm dd}\simeq-\frac{1}{4}\ln^2\xi$, and not Gaussian. The functions are log-normal mixtures of the KNO functions of one and of two BFKL cascades, with the moments of \eq{eq:moments}. The relation $\Psi_{\rm dd}=\sqrt{\Psi_{\rm BK}}$ is approached only at very large $\xi$. The constants in the entropy are $c_{\rm BK}=0.81$ and $c_{\rm dd}=0.09$.

(2) The KNO scaling, and with it the entropy $S_E=\ln\bar n$, holds only if the power-like factors $1/n$ and $1/n^2$ in the coefficients of the series are discarded. With these factors the mean multiplicity is not proportional to the single-Pomeron exchange.

(3) The contour representation of Ref.~\cite{Levin:2024ire} leads to functions that change sign (they are negative at $0.87<\xi<12.1$ for dipole-target scattering and at $0.22<\xi<1.9$ for dipole-dipole scattering) and cannot be interpreted as multiplicity distributions. The relation $\Psi_{\rm dd}\Lb\xi\Rb=\Psi_{\rm BK}\Lb\sqrt{\xi}\Rb$ is not supported.

(4) The sum of the series is not unique even within positive functions: the factor $\exp\Lb n^2/2\Rb$ defines an indeterminate moment problem. In the family of \eq{eq:family} the ambiguity is below $1\%$ at $\xi\leq10$ and grows in the tail.

(5) In the parton cascade with dipole merging the resummed series reproduces neither the exact amplitude at large rapidity nor the multiplicity distribution, because the coefficients of the series are used beyond the range in which they are known.

(6) In the UTM the resummed series reproduces the exact amplitude at all rapidities. The multiplicity distribution derived from it with the AGK cutting rules is the distribution of the number of inelastic interactions between two dilute cascades. It agrees with the exact distribution in the symmetric frame within about $20\%$ in the mean at $\gamma\langle n\rangle_{\tilde Y/2}\lesssim0.1$, deviates from it by a factor of about $3$ at $\tilde Y=2\tilde Y_s$, and fails at larger rapidities, where the exact mean grows as a power of rapidity, the distribution has no KNO scaling, and the entropy grows as $\h\ln\langle k\rangle$ in the rest frame of the target and as $\frac{3}{4}\ln\langle k\rangle$ in the symmetric frame. The number of inelastic interactions is frame dependent.

(7) The amplitude and the multiplicity are determined by different configurations of the cascades: the amplitude by rare configurations with far fewer dipoles than the typical ones, whose probabilities are little affected by saturation, and the multiplicity by the typical ones. In the language of the multi-Pomeron series, all multi-Pomeron exchanges contribute to the mean number of dipoles, because the exact eigenfunctions vanish linearly, and not as the $k$-th power, at $u\to1$.

For QCD these results mean that the multiplicity distributions and the entropy obtained from the sum of large Pomeron loops rest on assumptions which the agreement of the amplitudes does not test: the behavior of the coefficients $C_n$ at small $n$, the choice of the measure in the summation, and the applicability of the dilute counting of cut Pomerons in the saturation region. A calculation of the multiplicity distribution which does not rely on these assumptions, for example from the evolution equations for the cross sections of fixed multiplicity in QCD~\cite{Kovchegov:1999ji,Levin:2024wtl}, would be needed to settle the form of the KNO function for dipole-dipole scattering. We hope that the explicit formulas and the exact benchmarks presented here will be useful for such a calculation.

\begin{acknowledgments}
The author thanks colleagues at UTFSM university for encouraging discussions. This research was supported by ANID grant Fondecyt \textnumero 3230699.
\end{acknowledgments}

\appendix

\section{Steepest descent with the Lambert function}
\label{app:lambert}

In this appendix we give the details of the evaluation of Eqs.~(\ref{eq:PsiBKn}) and (\ref{eq:Psiddn}) and the first corrections to Eqs.~(\ref{eq:PsiBKsd}) and (\ref{eq:Psiddsd}).

\subsection{Dipole-target scattering}

The exponent of \eq{eq:PsiBKn} and its derivatives are
\bea \label{eq:Af}
f\Lb\lambda\Rb &=& -\h\lambda^2+\lambda-\xi\, e^{\lambda}, \qquad f'\Lb\lambda\Rb = 1-\lambda-\xi\, e^{\lambda},\nn\\
f''\Lb\lambda\Rb &=& -1-\xi\, e^{\lambda},\nn\\
f^{(m)}\Lb\lambda\Rb &=& -\xi\, e^{\lambda}\quad \Lb m\geq3\Rb .
\eea
Since $f''<0$ for all real $\lambda$, the function $f$ is concave and has a single maximum. The condition $f'=0$ can be written as $\Lb 1-\lambda\Rb e^{1-\lambda} = e\,\xi$, whose solution is $1-\lambda_{SP} = W\Lb e\,\xi\Rb$, where $W\Lb x\Rb$ is the principal branch of the Lambert function, $W e^{W}=x$~\cite{Corless:1996zz}. At the saddle point $\xi e^{\lambda_{SP}} = W$, and
\bea \label{eq:Afsp}
f\Lb\lambda_{SP}\Rb &=& \h - W - \h W^2,\nn\\
f''\Lb\lambda_{SP}\Rb &=& -\Lb 1+W\Rb,\nn\\
f^{(m)}\Lb\lambda_{SP}\Rb &=& -W \quad \Lb m\geq3\Rb .
\eea
The Gaussian integration around the saddle point gives \eq{eq:PsiBKsd}. The first correction follows from the standard expansion of the Laplace method:\verificado
\bea \label{eq:ANLO}
\Psi_{\rm BK}\Lb\xi\Rb &=& \frac{e^{\h - W - \h W^2}}{\sqrt{1+W}}\Lb 1+\delta_{\rm BK}+\dots\Rb ,\nn\\
\delta_{\rm BK} &=& \frac{5\,W^2}{24\Lb 1+W\Rb^3}-\frac{W}{8\Lb 1+W\Rb^2} .
\eea
The correction vanishes at $\xi\to0$, where \eq{eq:PsiBKsd} is exact, and decreases as $1/(12\,W)$ at $\xi\to\infty$; its maximal absolute value is $0.014$, at $W=0.32$. The steepest-descent result is therefore uniformly accurate, and not only an asymptotic estimate. With $W\Lb x\Rb = \ln x-\ln\ln x+\mathcal{O}\Lb\ln\ln x/\ln x\Rb$ at large $x$ one obtains \eq{eq:tailBK}.

\subsection{Dipole-dipole scattering}

With $K_0\Lb x\Rb = \sqrt{\pi/(2x)}\,e^{-x}\Lb 1-1/(8x)+\dots\Rb$ and $x=2\sqrt{\xi}\,e^{-\lambda/2}$, \eq{eq:Psiddn} takes the form
\bea \label{eq:Ag}
\Psi_{\rm dd}\Lb\xi\Rb &\simeq& \frac{\xi^{-1/4}}{2}\intl^{\infty}_{-\infty} d\lambda\; e^{g\Lb\lambda\Rb},\nn\\
g\Lb\lambda\Rb &=& -\frac{\lambda^2}{4}-\frac{3\lambda}{4}-2\sqrt{\xi}\,e^{-\lambda/2} .
\eea
Denoting $\zeta\equiv\sqrt{\xi}\,e^{-\lambda_{SP}/2}$, the condition $g'=0$ reads $\lambda_{SP}=2\zeta-\frac{3}{2}$, and therefore $\zeta\,e^{\zeta}=e^{3/4}\sqrt{\xi}$, which is \eq{eq:spdd}. At the saddle point
\bea \label{eq:Agsp}
g\Lb\lambda_{SP}\Rb &=& -\zeta^2-2\zeta+\frac{9}{16}, \qquad g''\Lb\lambda_{SP}\Rb = -\h\Lb 1+\zeta\Rb,\nn\\
g'''\Lb\lambda_{SP}\Rb &=& \frac{\zeta}{4}, \qquad g''''\Lb\lambda_{SP}\Rb = -\frac{\zeta}{8} ,
\eea
which gives \eq{eq:Psiddsd}, and the first correction is\verificado
\beq \label{eq:ANLOdd}
\delta_{\rm dd} \,=\, \frac{5\,\zeta^2}{48\Lb 1+\zeta\Rb^3}-\frac{\zeta}{16\Lb 1+\zeta\Rb^2}-\frac{1}{16\,\zeta} ,
\eeq
where the last term comes from the correction to the asymptotic form of $K_0$. In contrast to the dipole-target case, \eq{eq:Psiddsd} is not valid at small $\xi$, where $\zeta\to0$ and the asymptotic form of $K_0$ cannot be used.

Table~\ref{tab:nlo} compares the two orders of the expansion with a numerical integration performed with 30 significant digits. With the corrections of Eqs.~(\ref{eq:ANLO}) and (\ref{eq:ANLOdd}) the accuracy is about $10^{-3}$ in $\ln\Psi_{\rm BK}$ for all $\xi$ and better than $10^{-2}$ in $\ln\Psi_{\rm dd}$ for $\xi\geq1$.

\begin{table}[tb]
\caption{\label{tab:nlo} Logarithms of the KNO functions: numerical integration, steepest descent [Eqs.~(\ref{eq:PsiBKsd}) and (\ref{eq:Psiddsd})], and steepest descent with the first correction [Eqs.~(\ref{eq:ANLO}) and (\ref{eq:ANLOdd})].}
\begin{ruledtabular}
\begin{tabular}{rrrrrrr}
 & \multicolumn{3}{c}{$\ln\Psi_{\rm BK}$} & \multicolumn{3}{c}{$\ln\Psi_{\rm dd}$}\\
$\xi$ & exact & LO & NLO & exact & LO & NLO \\
\hline
$10^{-2}$ & $0.457$ & $0.460$ & $0.457$ & $1.611$ & $1.819$ & $1.375$\\
$10^{-1}$ & $0.147$ & $0.159$ & $0.146$ & $0.356$ & $0.475$ & $0.311$\\
$1$ & $-1.352$ & $-1.347$ & $-1.352$ & $-1.779$ & $-1.711$ & $-1.789$\\
$10$ & $-5.456$ & $-5.460$ & $-5.455$ & $-5.177$ & $-5.136$ & $-5.179$\\
$10^{2}$ & $-13.210$ & $-13.217$ & $-13.210$ & $-10.156$ & $-10.130$ & $-10.156$\\
$10^{4}$ & $-41.654$ & $-41.660$ & $-41.654$ & $-25.735$ & $-25.722$ & $-25.734$\\
$10^{6}$ & $-88.807$ & $-88.812$ & $-88.807$ & $-49.699$ & $-49.691$ & $-49.698$\\
\end{tabular}
\end{ruledtabular}
\end{table}

\subsection{Behavior at small $\xi$}

At $\xi\to0$, \eq{eq:PsiBKn} gives $\Psi_{\rm BK}\Lb\xi\Rb = e^{1/2}-e^{2}\,\xi+\mathcal{O}\Lb\xi^2\Rb$, which follows from the expansion of $\exp\Lb-\xi e^{\lambda}\Rb$ under the integral. For \eq{eq:Psiddn}, $K_0\Lb x\Rb\simeq-\ln\Lb x/2\Rb-\gamma_E$ at small $x$, where $\gamma_E$ is the Euler constant, and
\beq \label{eq:Asmall}
\Psi_{\rm dd}\Lb\xi\Rb \,\simeq\, e\Lb \ln\frac{1}{\xi}-2\gamma_E-2\Rb , \qquad \xi\ll1 .
\eeq
The singularity is integrable. It reflects the large probability to have a small number of interacting pairs when one of the two cascades contains few dipoles.

\section{The function $F\Lb\xi\Rb$ of the contour representation}
\label{app:contour}

Here we derive \eq{eq:Fstable} and the positions of the zeros of $F\Lb\xi\Rb$. In \eq{eq:Psiold} the integrand $e^{a\omega^2}\xi^{\omega}/\Gamma\Lb\omega\Rb$, $a=\bg^2\kappa/2$, is an entire function of $\omega$ which decreases as $\exp\Lb-a\,({\rm Im}\,\omega)^2\Rb$ along any vertical line, so that the contour can be shifted to ${\rm Re}\,\omega=c$ with arbitrary real $c$. With $\omega=c+it$,
\beq \label{eq:Bexp}
a\,\omega^2+\omega\ln\xi = a\,c^2+c\ln\xi-a\,t^2+i\,t\Lb 2ac+\ln\xi\Rb .
\eeq
The choice $c=-\ln\xi/(2a)$ removes the oscillating factor and gives $a c^2+c\ln\xi=-\ln^2\xi/(4a)$. Since $1/\Gamma\Lb c-it\Rb$ is the complex conjugate of $1/\Gamma\Lb c+it\Rb$, the imaginary part of the integrand is odd in $t$, and we arrive at \eq{eq:Fstable}. The remaining integral converges as a Gaussian and has no large cancellations for any $\xi$, in contrast to the integral along a fixed contour, in which $\xi^{\omega}$ oscillates. For this reason the direct numerical evaluation along ${\rm Re}\,\omega=\epsilon$ loses accuracy both at $\xi\ll1$ and at $\xi\gg1$.

For $\xi>1$ one has $c<0$. Using the reflection formula,
\beq \label{eq:Breflect}
\frac{1}{\Gamma\Lb c+it\Rb} \,=\, \frac{1}{\pi}\,\Gamma\Lb 1-c-it\Rb\,\sin\Lb\pi\Lb c+it\Rb\Rb ,
\eeq
one sees that the integrand of $R\Lb\xi\Rb$ is dominated by a term proportional to $\sin\Lb\pi c\Rb\cosh\Lb\pi t\Rb$, which changes sign each time $c$ decreases by unity, with an amplitude that grows as $\Gamma\Lb1-c\Rb$. At large $|c|$ the zeros of $F$ are therefore equally spaced in $c$, i.e., the ratio of consecutive zeros in $\xi$ tends to $e^{2a}$. For $a=0.96$ the ratios of the first zeros of \eq{eq:zeros} are $13.8$, $10.5$, $9.3$ and $8.7$, slowly approaching $e^{2a}=6.8$. For $\xi<1$, $c>0$ and the integrand is dominated by the region of small $t$, in which ${\rm Re}\left[1/\Gamma\Lb c+it\Rb\right]>0$; the first zero, at $\xi=0.874$, appears at a small positive value of $c$, $c=0.07$.

The factor $e^{-\xi}$ in \eq{eq:Psiold} does not affect the sign. The function $\Psi^{(\omega)}_{\rm BK}=C e^{-\xi}F\Lb\xi\Rb$ is thus an oscillating function of decreasing amplitude.

\section{Exact relations in the zero-dimensional models}
\label{app:exact}

\subsection{Generating function of the inelastic interactions}

Let $P_n$ be the probabilities to find $n$ dipoles in the projectile and let each dipole interact inelastically with the target with the probability $p$, independently of the others. The generating function of the distribution of \eq{eq:cuts} is
\beq \label{eq:genrest}
\sum_{k\geq0}\sigma_k\,x^k \,=\,\sum_n P_n\Lb 1-p+p\,x\Rb^n \,=\, Z\Lb 1-p\Lb1-x\Rb\Rb .
\eeq
In particular $\sigma_0=Z\Lb1-p\Rb$, $\sum_{k\geq1}\sigma_k=1-Z\Lb1-p\Rb$, and $\sum_k k\,\sigma_k = p\,Z'\Lb1\Rb = p\,\langle n\rangle$. If each dipole can interact several times, with a Poisson distribution with mean $2\gamma$, the argument of $Z$ is $\exp\Lb-2\gamma\Lb1-x\Rb\Rb$ instead; $\sigma_0$ is the same, and the mean is multiplied by $2\gamma/p$. The distributions used in the main text are normalized to the sum over $k\geq1$.

For the BFKL cascade, $P_n=\Lb1/N\Rb\Lb1-1/N\Rb^{n-1}$ with $N=e^{Y}$, and $Z\Lb u\Rb = u/\Lb N-\Lb N-1\Rb u\Rb$. Equation~(\ref{eq:genrest}) then gives
\beq \label{eq:CBFKL}
\sigma_k \,=\,\Lb 1-\frac{1}{N}\Rb^{k-1}\frac{\Lb p N\Rb^k}{\Lb pN+1-p\Rb^{k+1}}, \qquad k\geq1 ,
\eeq
the expression quoted in Sec.~\ref{sec:kappa}.

\subsection{Stationary distribution of the cascade with merging}

For a stationary solution of \eq{eq:Zkappa} the right-hand side vanishes, which for $0<u<1$ gives $Z''=\Lb2/\kappa\Rb Z'$, i.e., $Z=A+B\,e^{2u/\kappa}$. The condition $Z\Lb1\Rb=1$ is the conservation of probability. The condition $Z\Lb0\Rb=0$ expresses the fact that the state without dipoles cannot be reached: in \eq{eq:master} the merging rate vanishes for $n=1$. These two conditions give \eq{eq:Zinf}, which corresponds to
\beq \label{eq:CPinf}
P_n\Lb Y\to\infty\Rb \,=\, \frac{1}{e^{2/\kappa}-1}\,\frac{\Lb 2/\kappa\Rb^n}{n!}, \qquad n\geq1 .
\eeq
We have checked that the numerical solution of \eq{eq:master} approaches \eq{eq:CPinf}; for $\kappa=0.03$ the sum over $n$ of the absolute values of the differences is $2\times10^{-2}$ at $Y=12$ and $2\times10^{-5}$ at $Y=20$. The mean number of cut dipoles is then $p\,\langle n\rangle/\sigma_{in}\simeq 2p/\kappa$, the asymptotic value shown in Fig.~\ref{fig:kappa}(b).

\subsection{Eigenfunctions of the UTM}

Multiplying \eq{eq:utm} by $u^n$ and summing over $n$ one obtains
\beq \label{eq:CZeq}
\frac{\partial Z\Lb u,\tilde Y\Rb}{\partial\tilde Y} \,=\, \Lb u-1\Rb\Big[ Z\Lb u,\tilde Y\Rb - Z\Lb e^{-\gamma}u,\tilde Y\Rb\Big] .
\eeq
For the functions $\Phi_k$ of \eq{eq:Zutm},
\beq
\Phi_k\Lb e^{-\gamma}u\Rb \,=\, e^{\gamma k}\,\Phi_k\Lb u\Rb\,\frac{e^{-\gamma k}-1/u}{1-1/u} ,
\eeq
and therefore
\beq \label{eq:Ceigen}
\Lb u-1\Rb\Big[\Phi_k\Lb u\Rb-\Phi_k\Lb e^{-\gamma}u\Rb\Big] \,=\,\Lb e^{\gamma k}-1\Rb\Phi_k\Lb u\Rb ,
\eeq
so that $\Phi_k\Lb u\Rb\,e^{\Delta_k\tilde Y}$ is a solution of \eq{eq:CZeq} with $\Delta_k=e^{\gamma k}-1$. The function $\Phi_k$ has simple zeros at $u=e^{\gamma l}$, $l=0,\dots,k-1$. Hence at the points $u=e^{\gamma m}$ the expansion of \eq{eq:Zutm} is a finite sum, and the initial condition $Z\Lb u,0\Rb=u$ at these points determines the coefficients recursively; the first ones are $C_0=1$ and $C_1=e^{\gamma}$, in agreement with $C_k=e^{\gamma k(k+1)/2}$~\cite{Kovner:2024tin}. At $u<1$ the series is asymptotic. Since only the factor with $l=0$ vanishes at $u=1$, the derivative at this point is the product of the remaining factors, which gives \eq{eq:nseries} and \eq{eq:Philinear}.

\subsection{Backward equation}

Equation~(\ref{eq:backward}) follows from the first step of the evolution. During the interval $d\tilde Y$ a single dipole emits a second one with the probability $\Lb1-e^{-\gamma}\Rb d\tilde Y$, after which the system evolves as a cascade which starts from two dipoles. Hence $Z^{(1)}\Lb u,\tilde Y+d\tilde Y\Rb = \Lb1-\Lb1-e^{-\gamma}\Rb d\tilde Y\Rb Z^{(1)}\Lb u,\tilde Y\Rb+\Lb1-e^{-\gamma}\Rb d\tilde Y\,Z^{(2)}\Lb u,\tilde Y\Rb$. In general,
\beq \label{eq:Cback}
\frac{\partial Z^{(m)}}{\partial\tilde Y} \,=\,\Lb 1-e^{-\gamma m}\Rb\Lb Z^{(m+1)} - Z^{(m)}\Rb .
\eeq
This hierarchy closes only if $Z^{(m)}=\Lb Z^{(1)}\Rb^m$, which requires emission rates proportional to $m$, as in the BFKL cascade.

\section{Numerical methods}
\label{app:numerics}

The master equations (\ref{eq:master}) and (\ref{eq:utm}) are integrated as systems of ordinary differential equations for $P_n$, $1\leq n\leq n_{\max}$, with an implicit variable-order method and relative tolerance $10^{-10}$. The cutoff $n_{\max}$ is chosen so that the probability at $n_{\max}$ is below $10^{-15}$: $n_{\max}=1500$ for \eq{eq:master} and $n_{\max}\simeq1.1\,\tilde Y+10/\gamma+400$ for \eq{eq:utm}. The conservation of the total probability and the frame independence of \eq{eq:Sframe} are used as checks; the latter holds to a relative accuracy better than $10^{-10}$ (checked for $\tilde Y_0=\tilde Y$, $\frac{3}{4}\tilde Y$ and $\h\tilde Y$ at $\gamma=0.025$, $\tilde Y=100$ and $400$, and at $\gamma=0.1$, $\tilde Y=50$ and $200$).

The distributions of Eqs.~(\ref{eq:cuts}) and (\ref{eq:cutsutm}) in the rest frame of the target are computed as binomial mixtures for all $k$. In the symmetric frame the number of pairs $n\,m$ reaches $10^{7}$; the pairs $\Lb n,m\Rb$ are grouped in $4000$ logarithmic bins of $n\,m$, and the binomial distribution, continued to noninteger $n\,m$ with $\Gamma$ functions, is evaluated on a grid in $k$ which contains all integers up to $200$ and is logarithmic above. The mean obtained in this way coincides with the exact expression of \eq{eq:meanexact} to four digits.

The integrals over $\tau$ in Eqs.~(\ref{eq:Sutm}) and (\ref{eq:sAGKutm}) are computed in the variable $\ln\tau$, and the ratio $x^n/\Lb1+x\Rb^{n+1}$ is evaluated as $\exp\Lb -n\ln\Lb1+1/x\Rb-\ln\Lb1+x\Rb\Rb$, which avoids the loss of accuracy at large $x$. The sum over $j$ is a discrete convolution in $\ln\tau+\gamma j$. The mean of \eq{eq:sAGKutm} is computed from the analytic expression of \eq{eq:meanAGK}, since at large $\tilde Y$ the distribution extends to values of $n$ which exceed any practical cutoff.

The integrals of Eqs.~(\ref{eq:PsiBKn}), (\ref{eq:Psiddn}) and (\ref{eq:Fstable}) are computed by adaptive quadrature with 30 significant digits on intervals centered at the saddle points. The zeros of \eq{eq:zeros} are found by bisection. The finite-$N$ sums of Tables~\ref{tab:factors} and \ref{tab:eq32} are performed term by term.

The analytic statements marked in the text as verified were also checked symbolically or by an independent numerical evaluation. The programs are available from the author upon request.

\end{document}